\documentclass[12pt,preprint]{aastex}
\input epsf

\shorttitle{Stellar Rotation Census of B Stars}
\shortauthors{Huang et al.}

\begin{document}

\received{}
\accepted{}

\title{A Stellar Rotation Census of B Stars: from ZAMS to TAMS}

\author{Wenjin Huang\altaffilmark{1}}

\affil{Department of Astronomy \\
University of Washington, Box 351580, Seattle, WA 98195-1580;\\
hwenjin@astro.washington.edu}

\author{D. R. Gies}

\affil{Center for High Angular Resolution Astronomy\\
Department of Physics and Astronomy \\
Georgia State University, P. O. Box 4106, Atlanta, GA  30302-4106;\\
gies@chara.gsu.edu}

\author{M. V. McSwain}

\affil{Department of Physics, Lehigh University\\
16 Memorial Drive East, Bethlehem, PA  18015;\\
mcswain@lehigh.edu}

\altaffiltext{1}{Visiting Astronomer, Kitt Peak National Observatory, 
National Optical Astronomy Observatory, which is operated by the 
Association of Universities for Research in Astronomy (AURA) under 
cooperative agreement with the National Science Foundation.}

\slugcomment{Accepted for publication in ApJ}



\begin{abstract}
Two recent observing campaigns provide us with moderate dispersion
spectra of more than 230 cluster and 370 field B stars.  Combining
them and the spectra of the B stars from our previous investigations     
($\sim$430 cluster and $\sim$100 field B stars) yields a large,
homogeneous sample for studying the rotational properties of
B stars.  We derive the projected rotational velocity $V\sin i$,
effective temperature, gravity, mass, and critical rotation speed
$V_{\rm crit}$ for each star.  We find that the average $V\sin i$ is
significantly lower among field stars because they are systematically
more evolved and spun down than their cluster counterparts.  
The rotational distribution functions of $V_{\rm eq}/V_{\rm crit}$
for the least evolved B stars show that lower mass B stars are born 
with a larger proportion of rapid rotators than higher mass B stars. 
However, the upper limit of $V_{\rm eq}/V_{\rm crit}$ that may 
separate normal B stars from emission line Be stars (where rotation 
promotes mass loss into a circumstellar disk) is smaller among the 
higher mass B stars. 
We compare the evolutionary trends of rotation (measured according to 
the polar gravity of the star) with recent models that treat internal mixing. 
The spin-down rates observed in the high mass subset ($\sim 9 M_\odot$)
agree with predictions, but the rates are larger for the low mass group 
($\sim 3 M_\odot$).  The faster spin down in the low mass B stars matches well
with the predictions based on conservation of angular momentum in 
individual spherical shells.  Our results suggest the fastest rotators
(that probably correspond to the emission line Be stars) are 
probably formed by evolutionary spin up (for the more massive stars)
and by mass transfer in binaries (for the full range of B star masses). 
\end{abstract}



\setcounter{footnote}{1}

\section{Introduction}                              

The intermediate mass B-type stars are excellent targets to 
explore the effects of stellar rotation on evolution because 
the consequences of rotation are predicted to be more important 
among more massive stars \citep{eks08} and because there are
large numbers of B-stars to study (compared to the more massive
O-type stars) in our region of the Galaxy.  
In several previous studies \citep{hua06a,hua06b}, we developed
spectroscopic diagnostic tools that carefully account for the 
rotational distortion and gravitational darkening associated 
with rapid rotation.  These studies of B stars in young clusters
demonstrated that the stars spin down with advancing evolutionary
state when the polar gravity $\log g_{\rm polar}$ is taken as 
an indicator of evolutionary status.  Moreover, we also noticed 
that the rate of spin down appears to be faster (relative to the 
main sequence lifetime) among lower mass stars.  However, the
details remain uncertain, because of the limited size of the sample and 
incomplete sample coverage.  There is a pressing need for observational
results on a larger sample to help resolve questions surrounding the
connection between rotation and birthplace density \citep{hua08,wol08}
and internal mixing \citep{eks08,dec09,hun09}. 

Since then, we embarked on major spectroscopic programs in 2006 
(on cluster B stars) and in 2008 (on field B stars).
We now have much more data and much better coverage of the B stars
near the main sequence (MS).  In this paper, we report on the
new spectra and their analysis.  We describe in \S2 the two
observing programs that significantly increase our sample size 
and make possible a statistical analysis of rotational trends. 
We outline in \S3 how we determine the key parameters
of individual stars, and we present the results in tabular form.  
In \S4, we compare the projected rotational velocities
of the cluster B stars with those of the field B stars and we argue that
a systematic difference in evolutionary state explains the differences
in their average properties.  We explore the mass dependence of the 
rotational properties of young MS stars in \S5, and we examine in \S6 the 
issue of the minimum rotation rate necessary for rapid rotators to 
eject gas into a circumstellar disk and appear as emission line Be stars. 
We make a critical comparison in \S7 of the observed and predicted 
changes in rotational properties with advancing evolutionary state,
and we discuss in \S8 the consequences of our work for models of the
formation processes of Be stars.  Our results are summarized in \S9. 


\section{Observations}                       

Our first run was devoted to B stars in open clusters, and we 
selected targets in young clusters in the northern sky from 
the WEBDA database\footnote{http://www.univie.ac.at/webda/}.
The names, ages, and number of B star targets of these clusters 
are given in Table~1.  Only two of the clusters (NGC~869 and 884)
were the subject of our previous survey \citep{hua06a}, and 
the new observations cover a different set of stars. 
We made the observations with the Wisconsin-Indiana-Yale-NOAO (WIYN)
3.5-m telescope at Kitt Peak National Observatory (KPNO) during
2006 October 4 -- 7 UT.  We used the Hydra multi-object spectrometer to
obtain moderate resolution spectra ($R\sim4600$, using grating 1200@28.7
in second order) in the blue region  ($4000 - 4600$\AA).  
A G7 BG-39 filter was installed to block contamination from
first order flux.  The fiber configuration for 
each cluster field generally consisted of 20 -- 40 science stars,
5 -- 6 guide stars, and 10 -- 20 random sky positions for sky background
subtraction.  Though we lost some time during the first two nights due
to bad weather, we were able to obtain spectra of all the cluster fields 
on two different nights.  We made a wavelength comparison exposure 
for each night and every field to secure an accurate wavelength
calibration and to measure radial velocities with care. 
The errors in the radial velocity measurements (made through 
cross-correlation of the observed and synthetic line profiles) 
are estimated to be $\sim$6.5 km~s$^{-1}$ by comparing the two-night
spectra of those cluster B stars that are not spectroscopic binaries.
After subtracting the bias level, we used the 
IRAF\footnote{http://iraf.noao.edu/} procedure DOHYDRA to do the
rest of data reduction (traced and binned the
spectrum, cleaned cosmic rays, calibrated wavelength, removed the sky
background, applied the 1-D flat) to obtain the final reduced spectra.

\placetable{tab1}      

The second run focused on field B stars, i.e., those with no 
obvious connection with a known star cluster.   
Our sample was selected from the Smithsonian Astrophysical
Observatory (SAO) Star Catalog based on their spectral types.
We avoided selecting any B star known to be associated with a cluster
(marked as ``star in cluster'' in SIMBAD).  We also checked the 
Deep Sky Survey\footnote{http://stdatu.stsci.edu/cgi-bin/dss\_form/}
field images around some of our field B stars fainter than $V=8$, and
we deleted those stars found in crowded, cluster-like fields.
We obtained the field B star spectroscopic observations with
the KPNO 2.1~m telescope and Goldcam spectrograph (with a
$3072\times1024$ CCD detector, T3KC) during 2008 November 14 -- 18
and 2008 December 9 -- 12.  The spectrograph grating (G47 with 831
lines mm$^{-1}$) was used in second order with a CuSO$_4$ blocking
filter, and this arrangement provided spectral 
coverage of about 900\AA\ around a central wavelength of 4400\AA.  
The slit width was set at $1\farcs3$, leading to a resolving
power of $R\sim 2400$ (FWHM$\sim$1.83 \AA, measured in the comparison spectra).
The integration time for each exposure was set to reach a S/N ratio 
$> 200$ in the continuum regions.  We made comparison (HeNeAr lamp) exposures
for each stellar exposure to ensure accurate wavelength calibration of the spectra.
The errors associated with the wavelength calibration are approximately 
6 km~s$^{-1}$ based upon residuals of the comparison spectra fit and 
a comparison of multi-night measurements of radial velocities of 
stars that are not spectroscopic binaries.  We obtained the final 
reduced spectra using standard IRAF CCD image reduction methods  
(subtracted the bias level, divided the flat images, removed cosmic
rays, and fixed the bad pixels/columns) and long-slit stellar
spectrum extraction procedures (traced and binned the spectrum,
calibrated wavelength, and normalized to the continuum).


\section{Stellar Parameters}       

\begin{figure}
\plotone{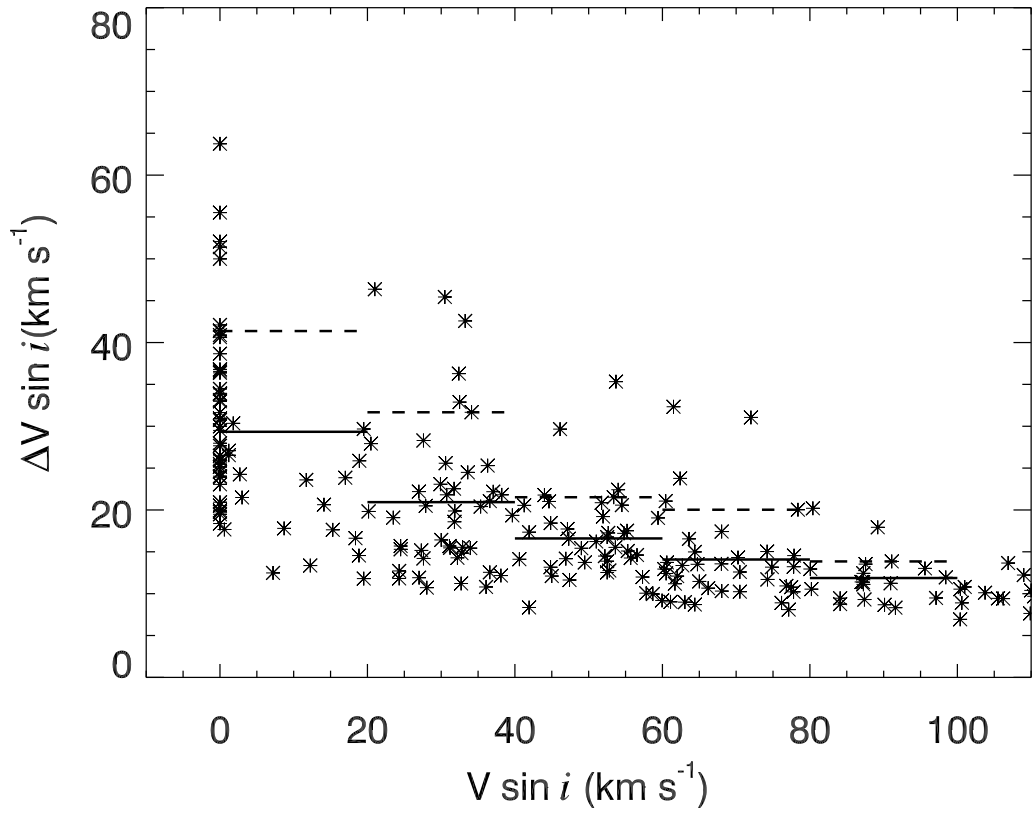}
\caption{A plot of $V \sin i$ versus its fitting error in the
low $V \sin i$ regime for our 2008 field B star sample.  The solid
lines indicate the mean value of $\Delta V \sin i$ of the stars in
each 20 km~s$^{-1}$ bin.  The dashed lines mark the positions that
are larger than the $\Delta V \sin i$ values of 90\% of the stars
in each bin.}
\label{fig0}
\end{figure}

The procedure to determine the physical parameters of 
B stars is similar to that used in our previous
studies \citep{hua06a,hua06b,hua08}.  First we measure
the $V \sin i$ values by fitting synthetic model profiles of \ion{He}{1}
$\lambda 4471$ and \ion{Mg}{2} $\lambda 4481$, using realistic
physical models of rotating stars (considering Roche geometry
and gravity darkening).  The details of this step are described
in \citet{hua06a}.  For narrow line spectra, the profile
fitting method may yield $V\sin i$ values that are lower than
the spectral resolution.  In the low $V \sin i$ regime, the
shape change in a line profile due to rotational broadening
is less significant.  Therefore, the fitting errors
increase as the $V \sin i$ values drop below the spectral
resolution.  This trend is clearly shown in the $V \sin i$
measurements of our field star sample (see Fig.~\ref{fig0}).
For most of the field stars with $V \sin i < 40$ km~s$^{-1}$,
the fitting errors are comparable to or even larger than the
derived $V \sin i$ values.  For the stars with $V \sin i > 40$
km~s$^{-1}$, their $V \sin i$ values are significantly higher
than the fitting errors.  Thus, we use 40 km~s$^{-1}$ as the
lower limit of the reliable $V \sin i$ values that our
profile fitting method can measure for our 2008 field sample.
The spectral resolution of the other B star spectra we obtained
are similar to or better than that of the 2008 field sample.  Thus,
the lower limit of reliable $V \sin i$ values for them is similar
to or lower than 40 km~s$^{-1}$.

\placefigure{fig0}     

We then derive both the effective temperature ($T_{\rm eff}$)
and gravity ($\log g$) averaged over the visible hemisphere of
a rotating star by fitting the H$\gamma$ profile (see details in
\citealt{hua06b}).  We note that the synthesized H$\gamma$
profiles were created by the spectrum synthesis code,
SYNSPEC43\footnote{http://nova.astro.umd.edu/Synspec43/synspec.html},
using Kurucz LTE model atmospheres.  This ensures that the
stellar parameters of the new B targets are derived in the
same way as done in our previous surveys.  
More sophisticated non-LTE, line blanketed models yield 
model H$\gamma$ profiles that are very similar to those 
derived from Kurucz LTE model atmospheres (especially for MS stars).  
For example, \citet{lan07} show that the non-LTE H$\gamma$ profiles
tend to be slightly stronger and broader than those derived from 
LTE atmospheres (see their Fig.~7), which means that our derived
temperatures will slightly lower and gravities slightly higher 
than those from a non-LTE treatment.  These differences are 
insignificant for the analyses that follow in subsequent sections. 
The errors given in our final results (see Tables~3 and 4)
are numerical fitting errors (see \citealt{hua06a,hua06b} for the
derivation of these errors).  All the synthetic profiles were
convolved with the instrumental broadening profiles measured from the
corresponding comparison spectra before fitting to the observed
spectra.  By shifting the best fit profiles in wavelength
to match the observed profiles, we then obtained a radial velocity
for each night of observation.  The radial velocities were
transformed into the heliocentric frame by removing the orbital
motion of the Earth (derived using the RVCORRECT function in IRAF).  

Stellar rotation in a close binary system is significantly affected 
by strong gravitational interactions between the components. 
Consequently, their rotational evolution is dominated by tidal effects,
and we need to separate them from single stars
and long period binaries when we study stellar rotation.  We rely
on the radial velocity measurements to identify close binaries.
If the difference in the radial velocity measurements of a star
with multiple-night observations is larger than 13 km~s$^{-1}$
($\sim 2\sigma$), we classify it as a spectroscopic binary (SB). 
We refer to the rest as constant velocity (probable single)
stars.  To check the detection efficiency of SBs using
our 13 km~s$^{-1}$ criterion, we searched The Ninth
Catalogue of Spectroscopic Binary Orbits\footnote{http://sb9.astro.ulb.ac.be/}
(SB9; \citealt{pou04}) for known SBs among our field B star sample.  
We found a total of 20 systems in common that are listed in Table~2, and
we detected eight of these as SBs using the 13 km~s$^{-1}$ range limit. 
These are mainly short orbital period SBs, and the detection efficiency is 
approximately 70\% for periods $P < 14$ d.  Our SB-detection efficiency
quickly drops for SBs with longer orbital periods where the velocity 
changes over a 1 -- 2 d interval are small.  Note that 
although the constant velocity stars in our sample are
sometimes called single stars in later sections, some may be 
SBs with longer orbital periods whose rotationally-related properties
are probably similar to those of single stars. 

\placetable{tab2}      

The derived gravity ($\log g$) for a rotating star represents
an average of gravity over its visible hemisphere.  It may
not be a good indicator of evolutionary status of the star because
the effective gravity of the equatorial region is lowered by the 
centrifugal force induced by stellar rotation.  Instead, we use
the gravity at the poles of the star ($\log g_{\rm polar}$) to
estimate the age of the star since the $\log g_{\rm polar}$ value
is not significantly influenced by rotation.  The higher the $V \sin i$
value is,  the greater the difference between $\log g$ 
and $\log g_{\rm polar}$.  This difference,
$\Delta\log g$, also depends on stellar mass, radius, and
the inclination angle ($i$) of the spin axis.  Ideally, if we know
the inclination angle $i$, then we can determine $\Delta\log g$ from the
measured $V \sin i$, $T_{\rm eff}$, and $\log g$.  However, in practice,
$i$ is generally unknown.  Our strategy is to adopt a statistical
average value of $\Delta\log g$ over the range of possible inclination angle.
The detailed procedure is described in \citet{hua06b}.
Based on our model simulation results \citep{hua06b}, estimations
of $\log g_{\rm polar}$ are quite accurate for most situations (the
statistical errors are 0.03 dex or less).  For the extreme case where a
star is spinning very close to the breakup velocity
($V_{\rm eq}/V_{\rm crit} > 0.95$) and the inclination angle is low
($i < 20^\circ$), the estimated $\log g_{\rm polar}$ could be off by
$-0.1$ dex or worse, but such cases are rare.

\begin{figure}
\plotone{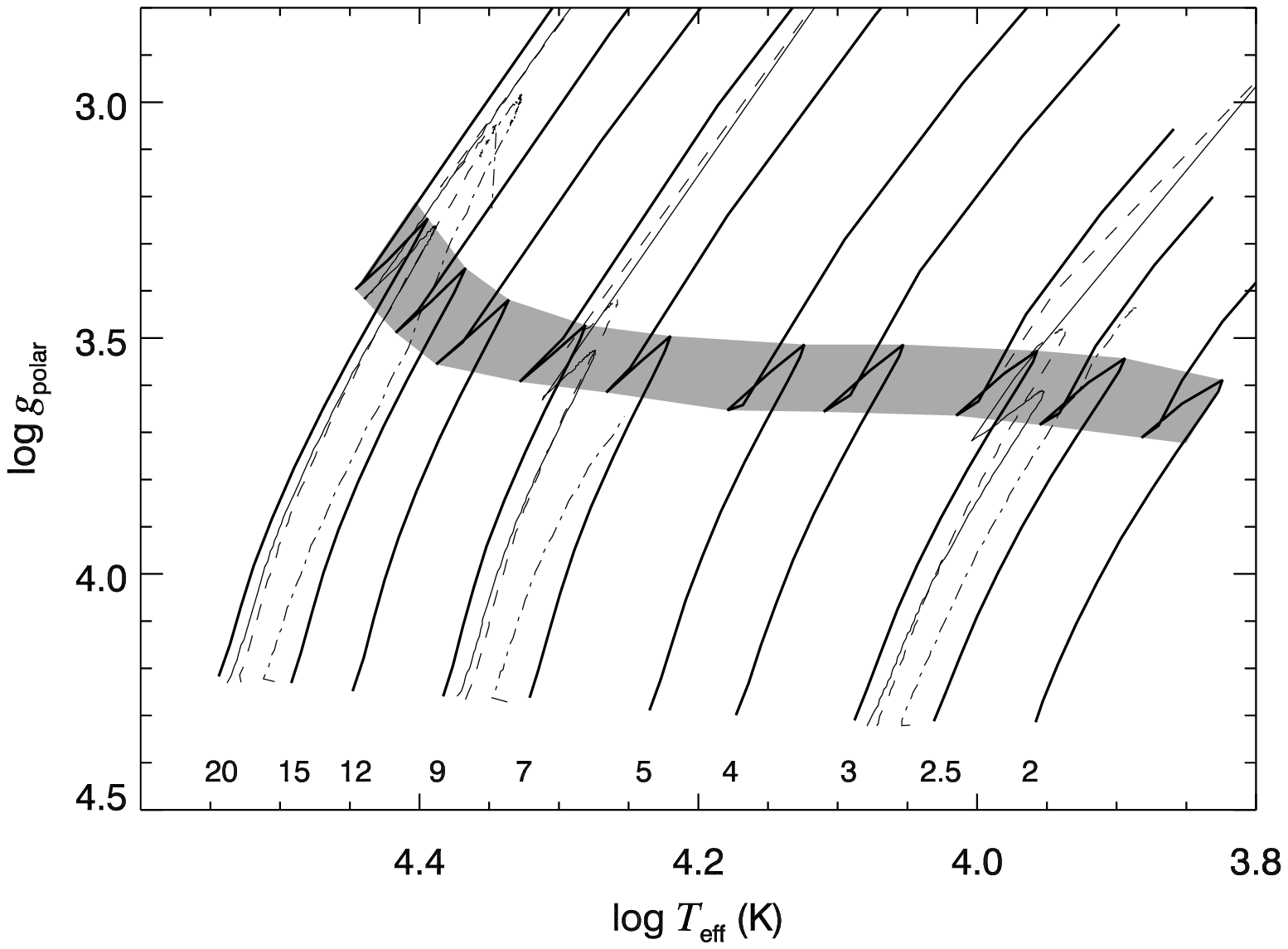}
\caption{A comparison of the evolutionary tracks for non-rotating stellar models
by \citet{sch92} ({\it thick solid lines}) and for rotating models by 
\citet{eks08} ({\it thin lines}: $\Omega/\Omega_{\rm crit}=0.1$; {\it dashed lines}: 
$\Omega/\Omega_{\rm crit}=0.5$; {\it dash-dotted lines}: $\Omega/\Omega_{\rm crit}=0.9$).
The masses of the rotating models are 20, 9, and 3 $M_\odot$ from left to
right, respectively.  The masses (in $M_\odot$) of non-rotating models
are marked by the numbers at the bottom.}
\label{fig1}
\end{figure}

We can approximately derive the stellar mass by assuming that the star 
follows a non-rotating star evolutionary track \citep{sch92} in the 
$(T_{\rm eff}, \log g)$ plane with a current position at 
$(T_{\rm eff}, \log g_{\rm polar}$). 
Here we assume that our derived hemisphere average $T_{\rm eff}$
is a reasonable choice for the whole surface average $T_{\rm eff}$.
Any differences between model tracks for rotating and non-rotating stars
will clearly influence the resulting mass estimate.  
We plot the evolutionary tracks for both non-rotating \citep{sch92} and
rotating \citep{eks08} models in Figure~1.  This comparison shows that the 
differences are minimal between the slow ($\Omega/\Omega_{\rm crit} =0.1$)
and moderately fast ($\Omega/\Omega_{\rm crit} =0.5$) rotating models, while
the evolutionary tracks of the very rapidly rotating models ($\Omega/
\Omega_{\rm crit} =0.9$) are shifted to lower temperature regions occupied 
by the lower mass, non-rotating models.  Our method may lead to a 
systematic underestimation of stellar mass by about 10\% for such rapid rotators.  
Note that there is a significant shift in effective temperature between the
non-rotating and slow rotating ($\Omega/\Omega_{\rm crit} =0.1$) tracks for
a given mass that indicates that there are systematic model differences
unrelated to rotation.  One other error source for the estimated mass is 
related to the placement in the ``zig-zag'' portion of the evolutionary tracks 
that corresponds to the terminal age main sequence (TAMS; gray area in Fig.~\ref{fig1}).  
Since stars spend more time on the cooler portion of the track compared to the 
hotter, advanced stage of evolution, we decided to use the low temperature
segment of the tracks in the shadowed area to determine stellar mass. The
derived mass values of those more evolved stars in this region could be
overestimated by up to 10\%, although the percentage of such stars is expected
to be low.

\placefigure{fig1}     

Once we estimate stellar mass, we can derive the polar radius ($r_{\rm p}$),
and then calculate the critical rotational velocity 
using the formula for the simple Roche model (point-source
gravitational plus rotational potential fields),
\begin{equation}
V_{\rm crit}=\sqrt{\frac{2}{3}\frac{GM}{r_{\rm p}}}.
\end{equation}
Here we can see how the mass errors propagate into $V_{\rm crit}$ errors. 
If mass were overestimated by 10\%, then $r_{\rm p}$ would be overestimated
by 5\% (from $\log g_{\rm polar}$), and $V_{\rm crit}$ would be overestimated
by only 2.5\%.  Thus, the errors in mass described in the previous
paragraph will result in relative errors in $V_{\rm crit}$ of 
$<2.5\%$.

The final derived parameters of our newly measured sample B stars are
given in Table~3 (cluster stars) and Table~4 (field stars) whose full contents
are available only in the electric form.  Table~3 lists the cluster name, 
stellar identification number from WEBDA, radial velocities, $V \sin i$, 
and the derived physical parameters.  The heliocentric Julian dates (HJD) 
of observation are given for each cluster in the electronic table header. 
Table~4 lists the star identifications in the SAO and Henry Draper (HD) 
catalogs, HJD of observation, and then the other quantities as in Table~3.   
The total numbers of B stars in our complete sample are summarized in 
Table~5 according to cluster or field membership and radial velocity variability status.  
Combining both new (first two rows in Table~5) and prior data (last two rows 
in Table~5) gives us a large and homogeneous sample ($\approx 1000$ stars) 
for the statistical analyses presented in following sections.

\placetable{tab3}      

\placetable{tab4}      

\placetable{tab5}      


\section{Field and Cluster Star Rotation Properties} 

We argued above that the rotational properties of binary stars 
may not be representative of B-stars as a whole, and so we set 
aside the subsample of radial velocity variable stars (probably binaries). 
The next issue to resolve is whether B stars in the field and 
in clusters have  such different rotational properties that 
the results for stars in the two categories need to be analyzed separately. 
Both \citet{abt02} and \citet{hua06a} found that nearby field B stars 
appear to rotate slower on average than B stars in young open clusters.  
The authors argue that this difference results from evolutionary
spin-down with time and the relatively greater age of the field stars.
The same systematic difference in rotation rates was also demonstrated in 
a series of papers by \citet{str05} and \citet{wol07,wol08}.  These authors argue instead 
that the trend results from the variation in density of the birthplace environment. 
To settle this issue, we recently presented an analysis of spectra of 108 
field B stars \citep{hua08} that was done in the same way as our earlier 
cluster study \citep{hua06b}.  This work showed that both the field 
and cluster stars exhibit a similar spin-down with advanced evolution 
(when the stars are grouped into common $\log g_{\rm polar}$ bins) and that field stars 
are generally more evolved than the those in clusters.  Here we repeat 
this comparison strategy with much larger samples.

\begin{figure}
\plotone{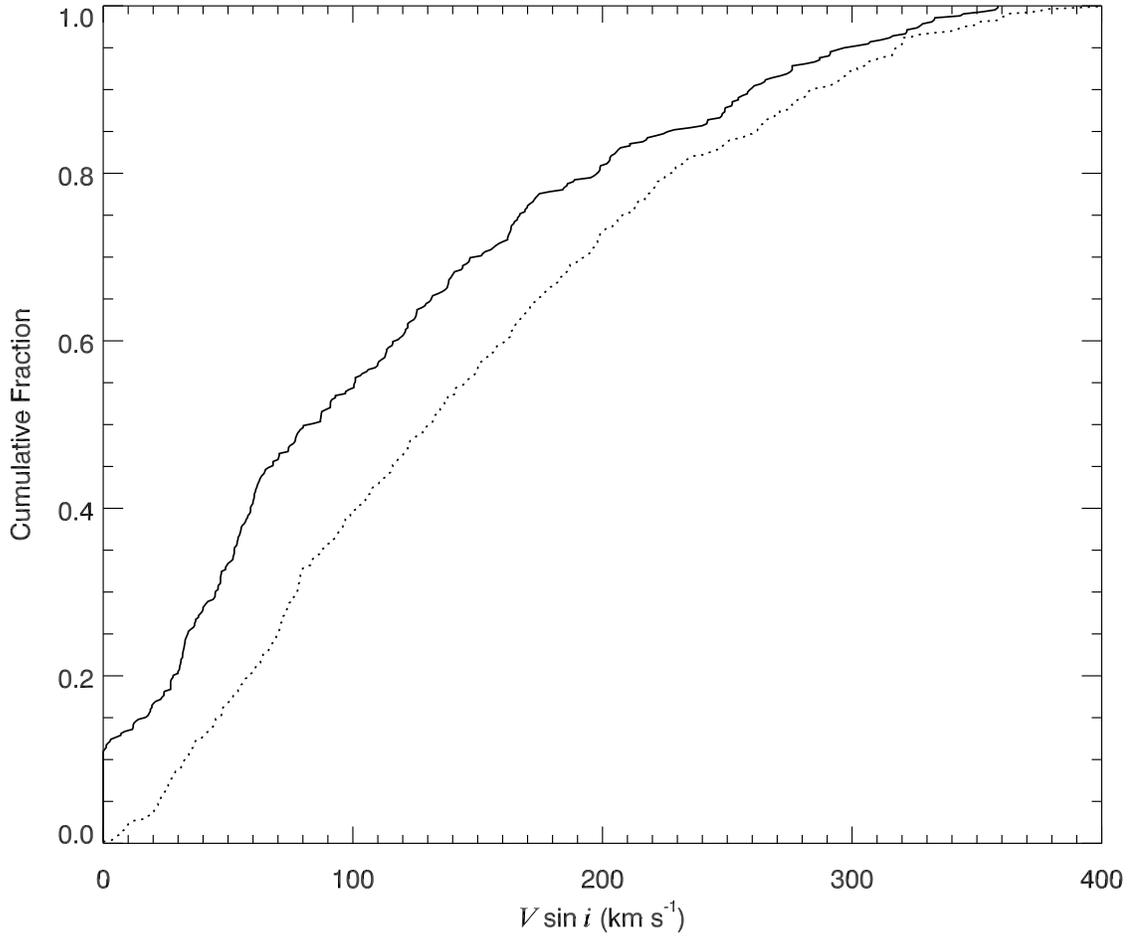}
\caption{Cumulative distribution functions of projected rotational velocity
for field ({\it solid line}) and cluster B stars ({\it dotted
line}).}
\label{fig2}
\end{figure}

\placefigure{fig2}     

The fact that, on average, cluster B stars rotate faster than 
field B stars is clearly confirmed again with our new B star
samples.  We show in Figure~\ref{fig2} the cumulative distribution functions of $V\sin i$ for 
the total samples of 419 field and 524 cluster stars (all constant velocity stars). 
The mean $V \sin i$ is $110.7\pm 4.6$ km~s$^{-1}$ for the field sample
and $145.2\pm 4.1$ km~s$^{-1}$ for the cluster sample.  A Kolmogorov-Smirnov (K-S) 
test shows that there is a very low probability ($2\times10^{-10}$)
that the two samples are drawn from the same population. 

\begin{figure}
\plotone{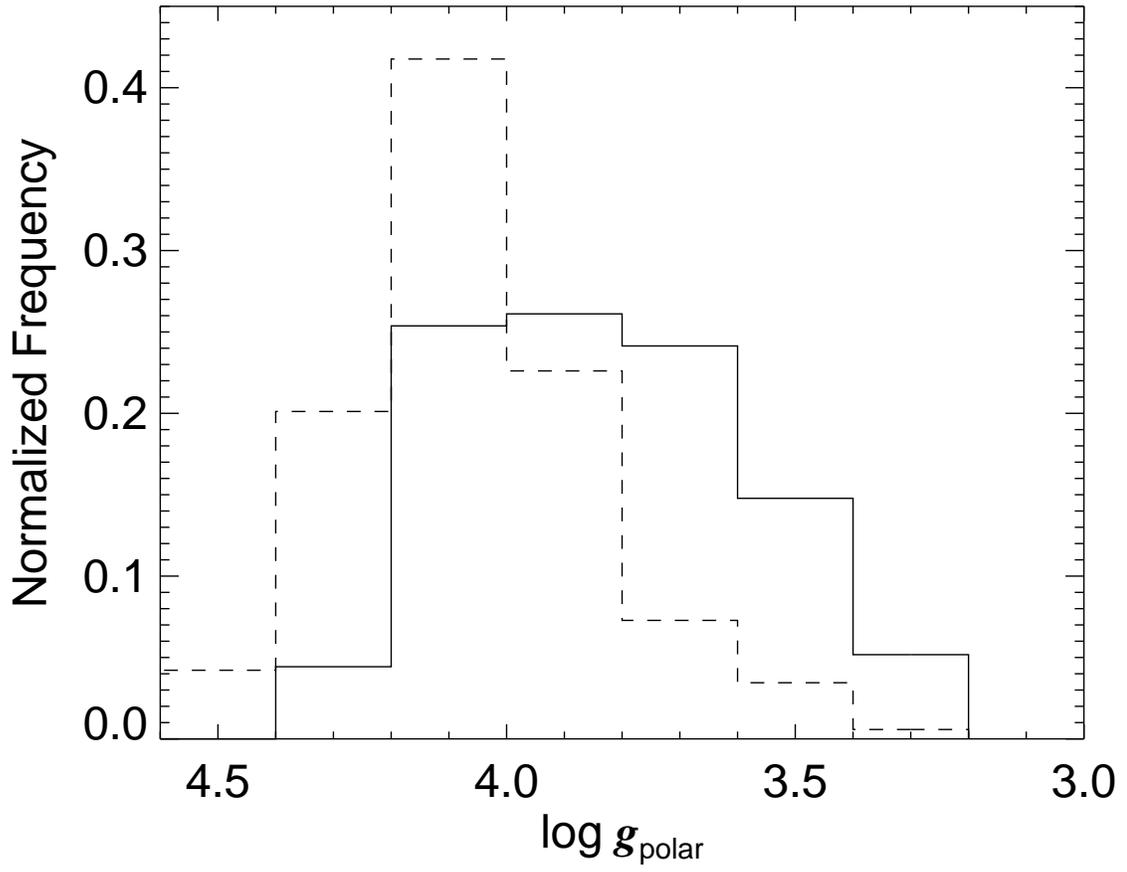}
\caption{The $\log g_{\rm polar}$ distributions of single field ({\it solid line})
and cluster ({\it dashed line}) B stars.}
\label{fig3}
\end{figure}

\placefigure{fig3}     

\begin{figure}
{\begin{center}
 \includegraphics[angle=0,height=17cm]{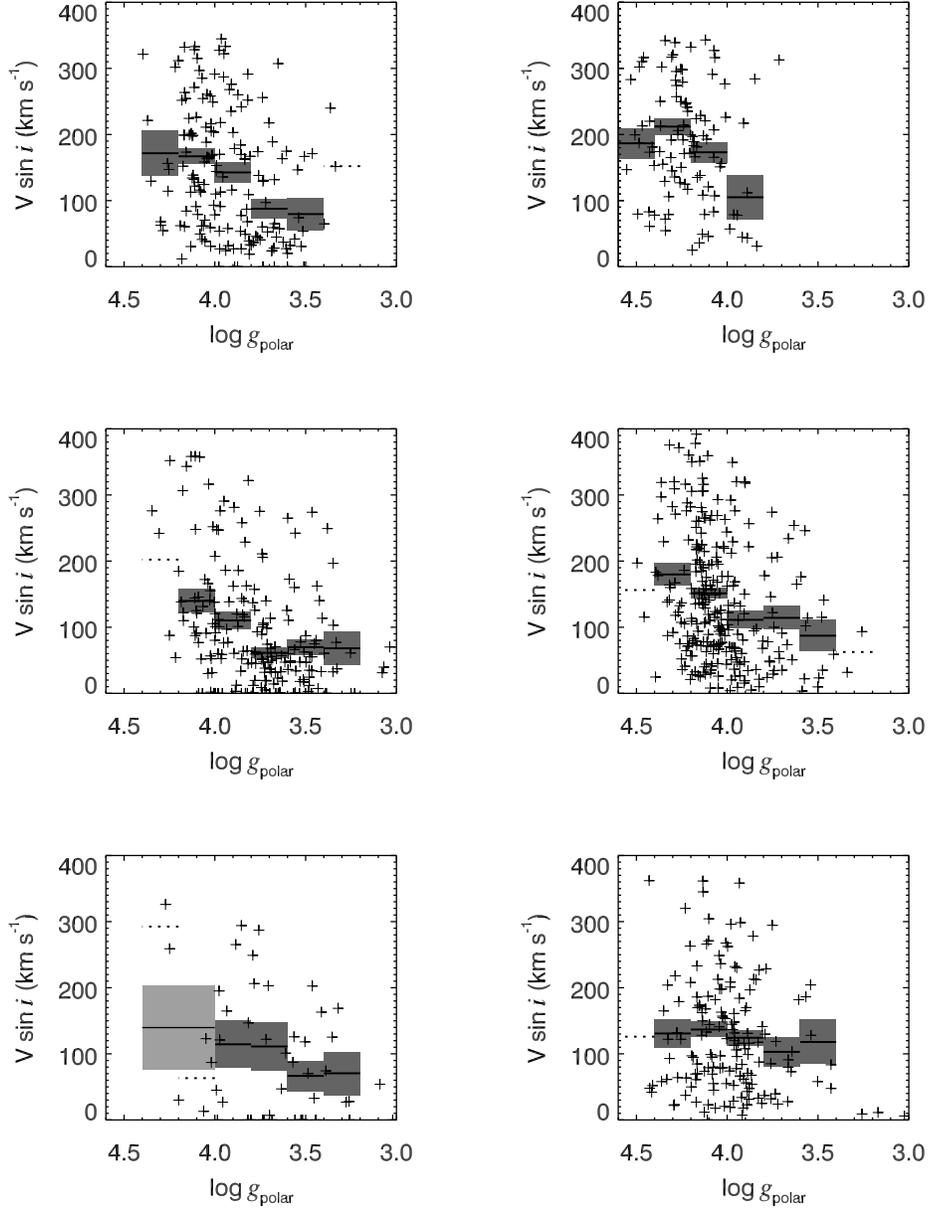}
 \end{center}}
\caption{Scatter plots of $V \sin i$ as a function of $\log g_{\rm polar}$ 
for the field ({\it left column}) and cluster samples ({\it right column})
and for three mass ranges: $2 < M/M_\odot < 4$ ({\it top}),
$4 \leq M /M_\odot < 8$ ({\it middle}), and $M/M_\odot \geq 8$ ({\it bottom}).  
The solid horizontal lines indicate the mean $V \sin i$ of each bin 
containing six or more stars while dotted lines indicate the same
for the bins containing fewer.  Shaded areas illustrate the 
standard deviation of the mean for each bin.
}
\label{fig4}
\end{figure}

If we plot the $\log g_{\rm polar}$ histogram for both the cluster and
field B stars (Fig.~\ref{fig3}), we can clearly see the difference in
evolutionary state of the two samples: the field B sample contains more
stars with lower $\log g_{\rm polar}$.  The mean $\log g_{\rm polar}$ is
3.80 for the field B star sample and is 4.11 for the cluster sample.
Therefore, we suggest that the slower rotation of our field sample is due solely to 
the greater influence of evolutionary spin down in a relatively older population. 
This can be demonstrated by constructing diagrams of mean $V \sin i$ 
as a function of $\log g_{\rm polar}$ for different mass groups among 
both the field and cluster B stars.  We show such plots in Figure~\ref{fig4} 
for the field ({\it left}) and cluster stars ({\it right}) for 
three mass groups: a low mass group ($2 < M/M_\odot < 4$; {\it top}), 
a moderate mass group ($4 < M/M_\odot < 8$; {\it middle}), and a high mass group
($M/M_\odot > 8$; {\it bottom}).  For each mass group, we calculate the mean of 
$V \sin i$ for stars that are binned according to $\log g_{\rm polar}$ 
(over a range of 0.2 dex for each bin).  Note that since 
there are only six young ($\log g_{\rm polar} > 4.0$) field stars in the high
mass subgroup, we consolidate two high $\log g_{\rm polar}$ bins into one as
shown in Figure~\ref{fig4}.  If the bin contains data for 
six or more stars, then we also calculate the standard deviation of the mean 
(shown as shaded areas in Figure~\ref{fig4}).  We find that the evolutionary
spin down trend is present in all subgroups of both the field and cluster samples
(although with shallower slope among the more massive stars). 
The results are also in qualitative agreement with those of
\citet{abt02} and \citet{abt03} who examined a similar number of B stars
and found that between luminosity class V (unevolved) and class III (evolved)
the mean projected rotational velocity declines by $\approx 15\%$ for early-B
types and by $\approx 40\%$ for late-B types.  If we compare the individual
bins of the field stars with the corresponding ones
of the cluster stars, we find that all of the matched pairs have similar
mean $V \sin i$ (within one standard deviation) except for one set with
$3.6 < \log g_{\rm polar} < 3.8$ in the mid-mass subgroup.  Note, however,
that the number of field B stars in the high mass group ($> 8 M_\odot$)
is small (a total of only 46 field B stars), and a larger sample would
help establish the details of the spin-down for this group.
These results are very similar to our previous findings \citep{hua08} which
were based on a smaller sample of field stars.  Thus, since the spin-down
trends appear to be similar in the field and cluster stars and since the
field stars tend to be more evolved (lower $\log g_{\rm polar}$), we again
conclude that the lower rotation rates of the field stars are primarily the
result of evolutionary spin-down changes.

In contrast to our results, \citet{wol07} did not detect a significant evolutionary
change in stellar rotation among their sample of stars, and they concluded
that differences in initial conditions and mass densities of star forming
regions are key to explaining the differences in rotational properties.
\citet{wol07} compared rotational properties of B0 -- B3 stars
(correponding to an approximate mass range of $6 - 12 M_\odot$) in young
and old clusters and associations in both low density ($\rho < 1 M_\odot ~{\rm
pc}^{-3}$) and high density ($\rho > 1 M_\odot ~{\rm pc}^{-3}$) environments.
They found that a small spin-down was evident when comparing the cumulative
distribution functions of $V\sin i$ for young and old dense clusters (see
their Fig.~3).  This relatively small spin-down rate probably results in
part because their sample consists mainly of more massive B-stars and
because the spin-down rates may be slower (as a function of $\log g_{\rm polar}$)
among massive B stars (see Fig.~5).  Furthermore, some of the spin-down trend
may be lost when samples are binned according to cluster age.  The time scale
of interest for evolutionary changes is the main sequence lifetime, which
depends sensitively on stellar mass, $\tau_{\rm MS} \propto M^{-2}$.
According to \citet{sch92}, a $6 M_\odot$ star has $\tau_{\rm MS} = 63$~Myr
while a $12 M_\odot$ star has $\tau_{\rm MS} = 16$ Myr.  Thus,
while the young age group of cluster stars (1 -- 6 Myr) considered by
Wolff et al.\ consists of mainly unevolved stars, the group of older cluster
stars (11 -- 16 Myr) probably includes some evolved massive stars
($69 - 100\% ~\tau_{\rm MS}$ for $M=12 M_\odot$) and many unevolved, lower
mass stars ($17 - 25\% ~\tau_{\rm MS}$ for $M=6 M_\odot$).  Thus, depending
on the mass distribution of stars, the rotational properties of the older
group may show little or no evidence of evolutionary spin-down because of
the mixture of evolutionary states represented within the sample.

\placetable{tab6}      

\citet{wol07} found much larger differences in the rotational
properties between stars in high and low density environments.  By their
criterion, all the cluster stars that we observed will belong to the high
density group (and all the field stars to the low density group).  We estimated
cluster densities (for both the clusters listed in Table~1 and those observed
earlier that are listed in Table~4 of Huang \& Gies 2006a) from the total
masses derived by Piskunov et al.\ (2008; their $\log M_{\rm A}$) and
from the cluster angular radii and distances from Kharchenko et
al.\ (2005; using their $R_{\rm core}$).  We have five clusters in
common with the sample from Wolff et al.: NGC 869, 884, 2244, 7380,
and IC 1805.  The density comparison is difficult for NGC 869 and 884
because these clusters overlap in the sky, and their $\log M_{\rm A}$
values are overestimated \citep{pis08}.  For NGC 2244, 7380,
and IC 1805, the derived mass densities are 5, 18, and 15 $M_\odot$ pc$^{-3}$,
respectively, which are comparable to the corresponding results of 11, 5,
and 10 $M_\odot$ pc$^{-3}$ derived by \citet{wol07} using somewhat
different assumptions.  The mass density range of our sample of clusters
runs from 1 to 78 $M_\odot$ pc$^{-3}$ (and higher for NGC 869 and 884), and,
thus, all the clusters we observed belong to the high density category
defined by \citet{wol07}.  However, the distributions of $V\sin i$ in our
high (cluster) and low density (field) samples show no evidence of systematic
differences when binned according to polar gravity (see Fig.~5).
To make as best a comparison as possible with the mass range considered by
Wolff et al., we selected a subset of stars in the mass range $6 - 12
M_\odot$ and determined the rotational properties as a function of polar gravity.
The results (summarized in Table~6) indicate that there are no significant
differences (within the statistical errors) between the field and cluster mean
values.  Our results do not rule out the role of environmental factors
(especially among the more massive B-stars where our sample contains fewer
targets), but given that there is no evidence of systematic differences
between the cluster and field stars within our sample, we will consider the
combined set of field and cluster stars together in the following sections.

\placefigure{fig4}     


\section{Young Main Sequence B Stars}  

When we study stellar rotation and its evolution, there is one key aspect
that we want to glean from observations.  That is how fast stars rotate
when they just start off their main sequence journey at the ZAMS.  
These initial rates are important because they show us how fast stars
rotate at one of the most important moments in their lifetime,  ZAMS,
representing both the starting point of the relatively simple (smoothly changing)
MS stage  and the end point of the extremely complicated pre-MS stage.
Accurate stellar rotation data at the ZAMS can help us to understand the role
of angular momentum in stellar evolution both before and after the ZAMS 
(cf.\ \citealt{wol04}).  Theoretically, the ZAMS only defines a moment when hydrogen
fusion is ignited in the stellar core (the temporal boundary between 
``protostar'' and ``star'').  
The evolutionary change of stellar rotation after ZAMS is expected to be
smooth and slow.   Thus, by studying the rotational properties of the least
evolved B stars in our single ($V_{\rm r}$ constant) star sample,
we should be able to obtain a reliable picture about how fast the ZAMS stars
rotate.  Since the ZAMS line of theoretical stellar models
in the mass range of B stars has basically the same surface gravity
($\log g \sim 4.2$; see Fig.~\ref{fig1}),  we decided to select all single B stars
with $\log g_{\rm polar} > 4.15$ to form a subsample that we will call the 
Young Main Sequence (YMS) stars.

\begin{figure}
\plotone{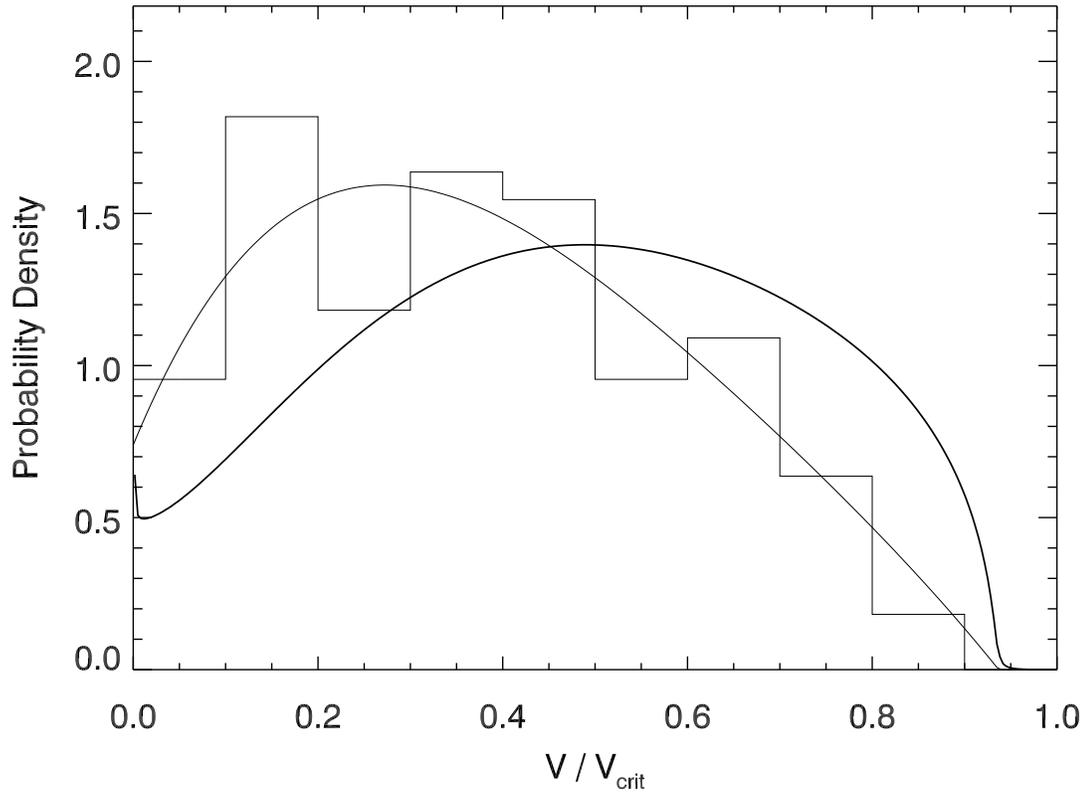}
\caption{The $V \sin i/ V_{\rm crit}$ histogram
of all young B stars in our sample with $\log g_{\rm polar} > 4.15$.
Its polynomial fit is plotted as a thin solid line.  The 
$V_{\rm eq}/ V_{\rm crit}$ distribution curve deconvolved from
the polynomial fit is plotted as a thick solid line.
}
\label{fig5}
\end{figure}

We have a total of 220 B stars in our YMS sample.  
The stellar rotation distribution of $V \sin i/V_{\rm crit}$ 
for the entire YMS sample is shown in Figure~\ref{fig5}.  
We made a simple polynomial fit of the histogram data 
of $V \sin i/V_{\rm crit}$ ({\it thin line}), and then  
we used the deconvolution algorithm from \citet{luc74} to derive the
distribution of $V_{\rm eq}/V_{\rm crit}$ ({\it thick line}).  Figure~\ref{fig5}
indicates that B stars in our sample are born with various rotation rates.
The highest probability density occurs around $V_{\rm eq}/V_{\rm crit} = 0.49$.
About 6\% of newborn B stars are very slow rotators
($V_{\rm eq}/V_{\rm crit} < 0.1$), while about 52\% of the B stars are born with
a value of $V_{\rm eq}/V_{\rm crit}$ between 0.4 and 0.8.  Only 1.3\% of B
stars are born as very rapid rotators ($V_{\rm eq}/V_{\rm crit} > 0.9$).  
\citet{wol04} have already shown from their data that the specific
angular momentum of very young B stars is spread well below the critical
limit.  Now we can see this more clearly in quantitative form in the 
$V_{\rm eq}/V_{\rm crit}$ distribution curve.

\placefigure{fig5}     

To investigate if these statistical data depend on stellar mass, 
we divided our YMS sample into three subgroups: low mass
($2 \leq M/M_\odot < 4$; 90 stars), middle mass ($4 \leq M/M_\odot < 8$; 92 stars), 
and high mass ($M/M_\odot \geq 8$; 38 stars).  We would expect to see 
distribution curves similar to that in Figure~\ref{fig5} if YMS B
stars in the different mass subgroups were born with a similar spin rate distribution.  
However, as shown in Figure~\ref{fig6}, the statistics clearly reveal a different
story: as stellar mass increases, more and more stars are born as
slow rotators (with lower $V_{\rm eq}/V_{\rm crit}$ values).  The fraction
of slow rotators (say, $V_{\rm eq}/V_{\rm crit} < 0.5$) in each subgroup are
37\%, 53\%, and 84\% for the low, middle, and high mass subgroup, respectively.

\placefigure{fig6}     

The statistical results on the stellar rotation rates of the
YMS B stars present some very interesting findings: 1) Even if we generally
accept that most massive stars are born as rapid rotators, there does
exist a significant fraction of newborn B stars that are very slow
rotators (e.g., $V_{\rm eq}/V_{\rm crit} < 0.1$).  There must exist
some processes that efficiently brake these slow rotators early on.
2) Compared to the less massive stars, the more massive stars tend
to be born as slow rotators.  Qualitatively, two factors may contribute
to this. First, we know that stronger stellar winds occur in more massive
stars, and these can effectively remove angular momentum \citep{mey00}.
Secondly, we know that the binary frequency is high among the
massive O-stars \citep{mas09}.  It may be that star formation
processes may tend to deposit angular momentum more in orbital motion
and less in stellar spin among the more massive stars \citep{lar07,lar09}.
\citet{abt09}, for example, found that stars in dense environments (usually
where more massive stars are found) tend to have more binary systems and
lower mean $V\sin i$ values.


\section{Rotational Upper Limits and Be Stars}    

As discussed by \citet{por03}, the ``Be'' phenomenon is associated
with a wide range of B stars whose spectra have exhibited emission features in
the Balmer line regions at some point in time.  We focus here on the 
so-called classical Be stars, non-supergiant B stars whose spectra 
generally show broad, photospheric, absorption lines that are indicative
of fast rotation.  To form a gaseous
disk around a star and generate the emission features, it is generally
thought that the star has to rotate very fast so that the centrifugal
force significantly cancels the gravitational force and makes 
it relatively easy to launch gas into orbit.  How fast can a B star
spin before it becomes a Be star?  This is a controversial topic
in the recent literature on Be stars, and the fact that there is 
still no widely accepted answer is partially due to the difficulty of obtaining 
reliable $V_{\rm eq}$ estimates for Be stars.  Because we can only 
derive $V \sin i$ from observational data, we require measurements 
of a large sample of Be stars to deal with the 
$\sin i$ factor in a statistical way.  \citet{yud01}
carried out a large rotation survey of Be stars and, by assigning
$V_{\rm crit}$ to each spectral subtype, he obtained values
of $<V_{\rm eq}>/<V_{\rm crit}>$ 
ranging from 0.5 to 0.8.  \citet{por96} investigated some Be-shell
stars which are presumably edge-on Be stars ($\sin i \sim 1$), and
his result suggests that the Be stars, if not different from
his Be-shell star sample, should rotate at 70$\sim$80\% of 
$V_{\rm crit}$.  This range was confirmed by the study of Be stars 
in the open cluster NGC~3766 by \citet{mcs08}, who compared the
observational $V \sin i$ cumulative distribution function with
the theoretical one.  However, \citet{tow04} point out that very rapidly
rotating stars may experience strong gravitational darkening  
towards their equatorial zones that leads to an underestimate
of the actual $V \sin i$ value.  They applied a correction for
this effect to the data available from \citet{cha01}, and their
results imply that Be stars are spinning close to the critical
rotation rate.  On the other hand, some later studies that took
the gravitational darkening effect into account \citep{fre05,cra05}
still favored relatively low $V_{\rm eq}/V_{\rm crit}$
values for Be stars.  \citet{how07} points out that the high
$V_{\rm eq}/V_{\rm crit}$ ratio (0.95 or higher) case only requires
``weak''  processes to lift the surface material into orbit, while
the low ratio case needs much more energy to send gas into orbit
(the velocity gap between surface and orbit can reach as much as
100 km~s$^{-1}$ in this case).
He suggests that, if Be stars can form in the low ratio situation,
the signatures of ``strong'' processes in stellar photosphere 
(such as  ``large-amplitude pulsations'') should be easily detected.
More recently, \citet{cra09} developed a detailed theoretical
pulsational model that suggests that a Keplerian disk could form
around Be stars with $V_{\rm eq}/V_{\rm crit}$ as low as 0.6.

Although our survey was focused primarily on the rotational properties of normal B stars,
we can use our sample to investigate this issue by considering the upper 
limit of $V_{\rm eq}/V_{\rm crit}$ for those stars that probably are
rotating at rates slightly below that required to form a Be disk.  
To achieve this goal, we first removed all the Be stars from our single 
($V_{\rm r}$ constant) star sample by excluding already known
Be stars and those with Balmer emission or deep, narrow, absorption lines 
(a signature of Be-shell stars) in our spectra.
This process led to a culling of 61 Be stars, leaving 
a total of 894 stars in the remaining sample (presumably all
non-emission line B stars).  These Be stars are identified in the
last column of Tables 3 and 4.  Note that our approach
here makes the tacit assumption that the processes leading to disk
formation require a certain minimum value of the ratio of equatorial
velocity to critical rotation.  Ideally, we could test this idea by
finding a minimum value of this ratio among the Be stars, but
unfortunately, it is difficult to apply our methods to Be stars
because of emission contamination in the profiles and the small
number of Be stars we observed.  Instead, we will simply compare
below our results on the fastest rotating, non-emission line
stars with results on the Be stars from other investigations.
Since the process of disk formation probably requires rapid
rotation and some other mechanical energy source
\citep[possibly pulsation;][]{riv03,cra09} and since the properties
of the latter may vary from star to star, we caution that the
$V_{\rm eq} / V_{\rm crit}$ criterion for disk formation will probably
apply only in some average sense.

We can approach the problem of finding a nominal upper limit on 
$V_{\rm eq}/V_{\rm crit}$ in two ways.  First, we can consider the 
observed upper limit in the distribution of $V \sin i / V_{\rm crit}$.  
One could simply adopt the largest derived value of this ratio 
as the upper limit for $V_{\rm eq}/V_{\rm crit}$, but because the errors 
associated with any given estimate are not negligible, we prefer to form a 
mean from a subset of stars with the largest observed ratios. 
Clearly, the larger the sample, the more well defined will be 
the upper limit boundary of the distribution.  
Those stars that help us set the upper limit will have a
$V_{\rm eq}/V_{\rm crit}$ ratio close to the limit and will have 
an inclination angle close to 90$^\circ$
(i.e., $\sin i \approx 1$).  If we choose the top 4\% of stars that
have the highest $V_{\rm eq}/V_{\rm crit}$ values (the fastest rotators)
and assume that their spin axes are randomly pointed in space, then about
$1/4$ of these stars will have $\sin i > 0.97$.  Thus, we expect that 
those stars in the top 1\% of the $V \sin i / V_{\rm crit}$ distribution 
probably correspond to very fast rotators (in the top 4\%) with
an edge-on orientation ($\sin i > 0.97$).  Therefore, for our sample of 
about 900 stars, the mean ratio from the top nine fast rotators of 
$V \sin i/V_{\rm crit} = 0.91$ provides an initial estimate of the 
limit separating the non-emission line and emission-line stars.
We list in the lower row of Table~7 this estimate plus the unique  
maximum and 3\% subsample estimates for the upper limit from all the 
$V \sin i / V_{\rm crit}$ measurements.

\placetable{tab7}      

\begin{figure}
{\begin{center}
 \includegraphics[angle=0,height=18cm]{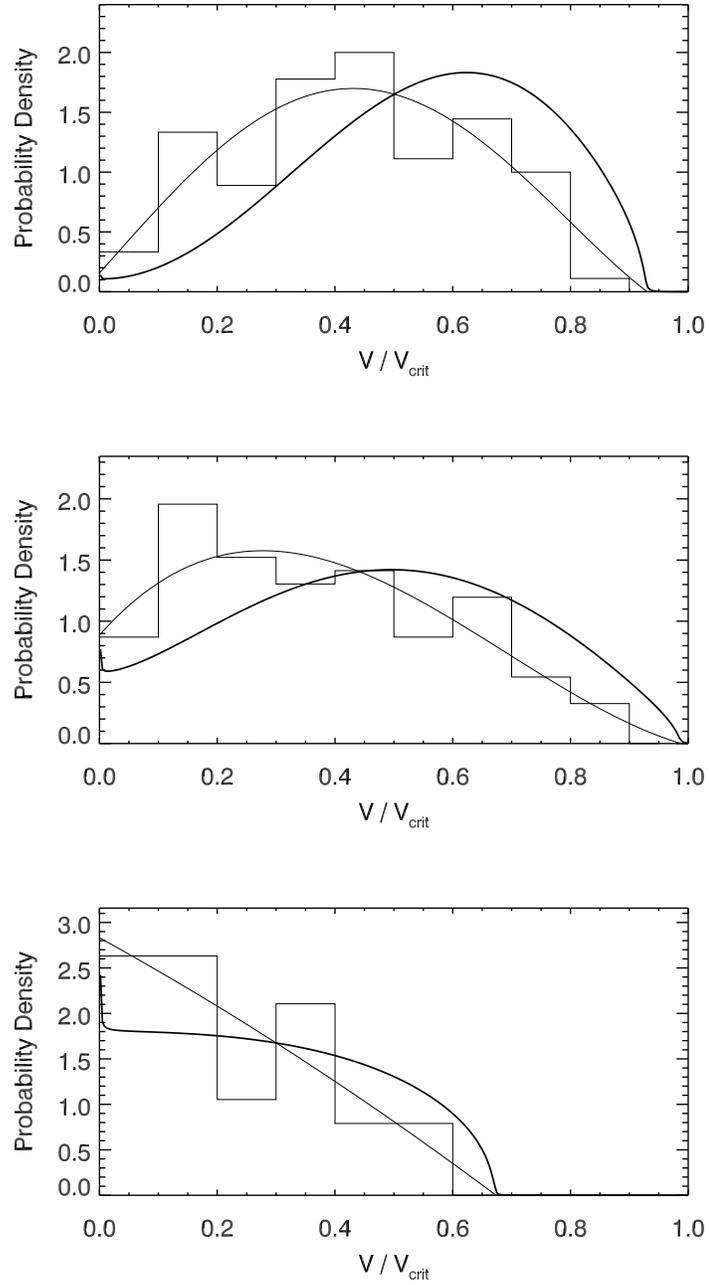}
 \end{center}}
\caption{Rotational velocity distributions (in the same format as Fig.~\ref{fig5}) 
for three subgroups of very young B stars with $\log g_{\rm polar} > 4.15$:
$2 < M/M_\odot < 4$ ({\it top panel}), 
$4 < M/M_\odot < 8$ ({\it middle panel}), and 
$M/M_\odot > 8$ ({\it bottom panel}).
}
\label{fig6}
\end{figure}

A second approach is to use the deconvolution method of \citet{luc74}
to derive the $V_{\rm eq}/V_{\rm crit}$ distribution from the 
$V \sin i / V_{\rm crit}$ distribution (as shown earlier in Fig.~\ref{fig5} and \ref{fig6}), 
and then determine the value of $V_{\rm eq}/V_{\rm crit}$ that marks 
the boundary for a given percentage of the distribution.  
The final row of Table~7 gives these boundaries for the top
4\% and 0.2\% of the reconstructed $V_{\rm eq}/V_{\rm crit}$ distribution
for the full sample of non-emission line stars.   The fact that 
the 4\% boundary ratio ($V_{\rm eq}/V_{\rm crit}=0.88$) is quite similar
to the estimate from the top 1\% of the  $V \sin i / V_{\rm crit}$ distribution 
($<V \sin i / V_{\rm crit}>_{\rm 1\%}=0.91$) indicates that both 
methods leads to similar and presumably reliable results. 

However, we found in \S5 that the rotational properties 
of the youngest B stars in our sample are strongly 
dependent upon the mass subgroup (slower among the more 
massive stars; see Fig.~\ref{fig6}), and thus, it is important 
to consider the upper limit of $V_{\rm eq}/V_{\rm crit}$ 
among samples of comparable mass.  In order to obtain meaningful
results from each mass subsample, we need to keep the population
large enough so that 1\% of their content comprises at
least 2 stars.  Thus, we divided our total of 894 stars into four
subsamples based upon mass.  The mass range of each subsample is given in
the first column of Table~7.  We made the same two statistical estimates
for the upper limit of $V_{\rm eq}/V_{\rm crit}$ for each subsample.
All the key statistical results are given in columns (2) to
(7) of Table~7.  The estimates based on the maximum value (from one star; column 3)
and the mean value of $<V \sin i/V_{\rm crit}>$ for the top 1\% of the stars
(based on 2--3 stars; column 4) suffer from small number
statistical errors.  Instead, estimates from $<V \sin i/V_{\rm crit}>$ of the
top 3\% of stars (5--8 stars; column 5) and from the boundary of $V_{\rm eq}/V_{\rm crit}$
for the top 4\% of the rapid rotators (6--10 stars; column 6) provide more reliable results. 

\begin{figure}
{\begin{center}
 \includegraphics[angle=0,height=18cm]{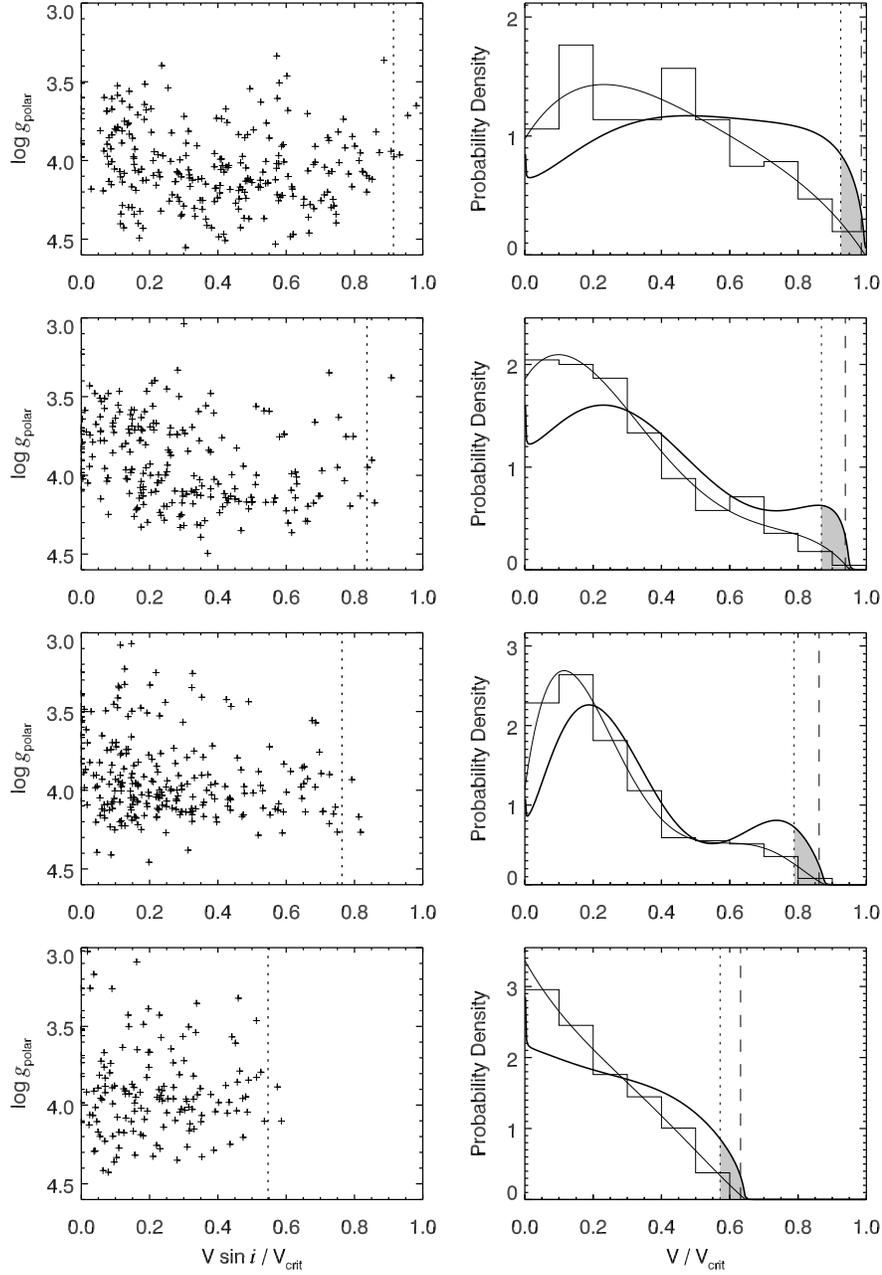}
 \end{center}}
\caption{Scatter plots of $(V\sin i /V_{\rm crit}, \log g_{\rm polar})$
and rotational distributions (in the same format as Fig.~\ref{fig6}) 
for non-emission line stars.  From top to bottom, 
each pair corresponds to mass ranges of 
$2.2 \leq M/M_\odot < 4.0$, 
$4.0 \leq M/M_\odot < 5.4$, 
$5.4 \leq M/M_\odot < 8.6$, and 
$M/M_\odot \geq 8.6$.  The shaded areas indicate the top 4\% stars
in the $V_{\rm eq} /V_{\rm crit}$ distributions. 
}
\label{fig7}
\end{figure}

We illustrate the rotational properties of the fastest spinning, 
non-emission line stars in two sets of panels in Figure~\ref{fig7}. 
The left hand panels show the distributions in the 
$(V \sin i / V_{\rm crit}$, $\log g_{\rm polar})$ plane with 
a dotted line marking the mean $<V \sin i/V_{\rm crit}>$ of the
top 3\% of stars, while the right hand panels show the 
histograms of $V \sin i/V_{\rm crit}$ and the deconvolved 
functions $V_{\rm eq}/V_{\rm crit}$ (format similar to Fig.~\ref{fig5} and \ref{fig6}).
The dotted and dashed lines in the right hand panels indicate 
the boundaries in the $V_{\rm eq}/V_{\rm crit}$ distribution
for the top 4\% and 0.2\% of the sample, respectively. 
These panels are arranged in four rows corresponding to the 
four mass subgroups (from low mass at the top to high mass at 
the bottom).  

\placefigure{fig7}     

We can immediately see from Figure~\ref{fig7} that the rotational upper limit 
for non-emission line B stars does indeed depend on stellar mass.  
Estimates of the upper limit from both $<V \sin i/V_{\rm crit}>_{\rm 3\%}$ 
and $(V_{\rm eq}/V_{\rm crit})_{\rm 4\%}$ show a consistent drop from about 
0.92 in the $3M_\odot$ subsample to about 0.56 in the $9M_\odot$ group.  
In principle, we might take the rotational upper limit as 
the value of $V_{\rm eq}/V_{\rm crit}$ where the deconvolved function 
reaches zero probability in Figure~\ref{fig7}.  However, the detailed shape of this 
function at the top end depends critically on the small numbers  
in the last few populated bins of $V \sin i/V_{\rm crit}$.  Thus,  
we arbitrarily adopted upper limit as the boundary of 
$V_{\rm eq}/V_{\rm crit}$ for the top 0.2\% of the 
deconvolved distribution, since this value is somewhat better 
constrained and corresponds to the case where we might find 0 or 1 star 
in our subsamples.  It is hard to estimate the error bars of the
adopted upper limit of $V_{\rm eq}/V_{\rm crit}$ from regular methods.  
However, since the $<V\sin i/V_{\rm crit}>_{3\%}$ and $(V_{\rm eq}/V_{\rm crit})_{4\%}$
statistical values in Table~7 are quite reliable and comparable, their mean
can can serve as the lower boundary to the upper limit of $V_{\rm eq}/V_{\rm crit}$. 
Then a working estimate of the error in the upper limit can be formed
by taking the difference between $(V_{\rm eq}/V_{\rm crit})_{\rm 0.2\%}$ 
and the mean of $<V\sin i/V_{\rm crit}>_{\rm 3\%}$ and 
$(V_{\rm eq}/V_{\rm crit})_{\rm 4\%}$.  Our final estimates of the 
rotational upper limits are given in column 7 of Table~7. 

These results suggest that the rotational threshold required to 
create a Be star disk varies dramatically through the B star mass range. 
For example, we find that low mass B stars ($M < 4 M_\odot$, or 
spectral subtype later than B6 for MS stars) may rotate extremely fast 
($V_{\rm eq}/V_{\rm crit} \sim 0.96$) without creating an outflowing disk.   
This may explain why the fast rotator, Regulus 
(below the threshold at $V_{\rm eq}/V_{\rm crit} = 0.86$; \citealt{mca05}),
has never exhibited any Balmer line emission.
For stellar masses $>8.6 M_\odot$ (spectral subtype B2 or earlier for MS stars),
the threshold drops to $V_{\rm eq}/V_{\rm crit} \sim 0.63$. 
This result may help resolve the controversy surrounding the 
massive Be star, Achernar ($\sim 7 M_\odot$).  Interferometric observations
of Achernar by \citet{dom03} showed that the star has an exceptionally large 
rotational oblateness with $R_{\rm eq}/R_{\rm polar}=1.56$, 
which is above the value expected (1.5) for a star rotating at the 
critical rate.  However, a recent  spectroscopic analysis by \citet{vin06} 
indicates that Achernar may rotate at $\Omega/\Omega_{\rm crit}\sim0.8$, 
which is equivalent to $V_{\rm eq}/V_{\rm crit}\sim0.65$ 
(cf.\ Fig.~10 in \citealt{eks08}).  Note that this rate 
is probably near to the implied threshold we find for 
disk formation among the massive B stars, so Achernar's rotation 
rate and status as a Be star are probably consistent with our results. 
  
The mass dependence of the rotation limit was generally ignored in
past with the notable exception of \citet{cra05} who reported a very 
similar $V_{\rm eq}/V_{\rm crit}$ dependence on 
spectral-type.  He applied Monte Carlo forward
modeling to a sample of 462 classical Be stars to test various
$V_{\rm eq}/V_{\rm crit}$ distribution functions against the observed
$V \sin i/V_{\rm crit}$ distribution.  His conclusion was that hot B
stars ($T_{\rm eff} \gtrsim 21000$K) tend to have a low threshold of
$V_{\rm eq}/V_{\rm crit}$ (``well below unity'') while cool B stars
($T_{\rm eff} \lesssim 21000$K) have an increasing threshold of
$V_{\rm eq}/V_{\rm crit}$ (up to unity) as ``$T_{\rm eff}$ decreases
to the end of the B spectral range''.  \citet{how07} raised
two concerns about these results: 1) they were based on
very heterogeneous data sources, and 2) $V_{\rm crit}$ was assigned 
to individual stars according to spectral type.
However, these concerns do not apply to our sample, and 
the overall good agreement between our result (based on normal B stars)
and Cranmer's (based on classical Be stars) convinces us that
the rotational limit for Be star formation is indeed mass dependent.

We should note that most of the B stars in our sample were retained
for this analysis based upon an examination of only a few spectra 
for Balmer line emission.  However, Be stars are intrinsically 
variable in nature \citep{por03}, and in some cases 
(``transient Be stars'') the disk emission may
even disappear at times (cf.\ \citealt{mcs09}).
It is possible that some of the fast rotators used in our 
analysis are transient Be stars that we observed during an 
emission-free epoch.   If such Be stars do exist among those
high $V\sin i/V_{\rm crit}$ stars that we used to determine the
rotational upper limit, then we may have overestimated the limits. 
We list in Table~8 all the stars with high $V\sin i/V_{\rm crit}$ 
that helped define our derived upper limits, and we recommend 
spectroscopic H$\alpha$ monitoring of these targets to verify 
their status as normal (non-emission line) B stars.  

\placetable{tab8}      


\section{Rotational Evolution of Main Sequence B Stars}    

Theoretical studies \citep{heg00,mey00} 
demonstrate that the evolutionary paths of rapidly rotating,
massive stars can be very different from those for non-rotating stars.
Rapid rotation can trigger strong interior mixing, 
extend the core hydrogen-burning lifetime, significantly alter the 
luminosity, and change the chemical composition of the outer
layers over time.  The most detailed models to 
date are presented in a seminal paper by \citet{eks08} (=EMMB), 
who show how the equatorial velocities and surface abundances 
change with time according to the initial rotation rate, 
metallicity, and mass.  We have already shown in \S4 (Fig.~\ref{fig4}) 
that the mean rotational properties do vary with $\log g_{\rm polar}$, 
our measurement of evolutionary state.  In this section, we re-sample  
our rotational data into two mass categories from the EMMB model grid, 
and we compare the observed changes with $\log g_{\rm polar}$ to 
predictions from the EMMB models that are based upon an 
initial rotational distribution derived from the youngest MS stars. 

Our goal is to compare our rotational results with the EMMB models 
for solar metallicity (a reasonable assumption for nearby, Galactic, B stars). 
The EMMB models are presented for masses of 3, 9, 20, and $60 M_\odot$, 
and since massive stars with $M > 15 M_\odot$ are sparse in our sample,
we are limited to making the comparison with the EMMB models for 
$3 M_\odot$ and $9 M_\odot$.  We selected stars with $2 \leq M/M_\odot < 4$
(262 stars; $<M> = 3.2 M_\odot$) for the $3 M_\odot$ group  
and those with $7 \leq M/M_\odot < 13$ (241 stars; $<M> = 9.2 M_\odot$) 
for the $9 M_\odot$ group.  All the known and suspected Be stars were 
omitted from these samples because we cannot derive reliable $T_{\rm eff}$ 
and $\log g$ parameters from the emission-blended H$\gamma$ profile.  For each
group, we first derived the rotational distribution curve (probability density vs.\ 
$V_{\rm eq}/V_{\rm crit}$) of the YMS stars ($\log g_{\rm polar} > 4.15$; 
90 stars in the 3 $M_\odot$ group, and 49 stars in the 9 $M_\odot$ group)
using the same method described in \S5.  These functions
describe the initial distribution of $V_{\rm eq}/V_{\rm crit}$.  
Then the EMMB theoretical models can be used to show how these 
distributions should change with time and, consequently, how 
$<V \sin i/V_{\rm crit}>$ should change with time.  This mean ratio 
can be directly compared to our observational results as a function
of $\log g_{\rm polar}$ to investigate evolutionary changes. 

\begin{figure}
{\begin{center}
 \includegraphics[angle=0,height=17cm]{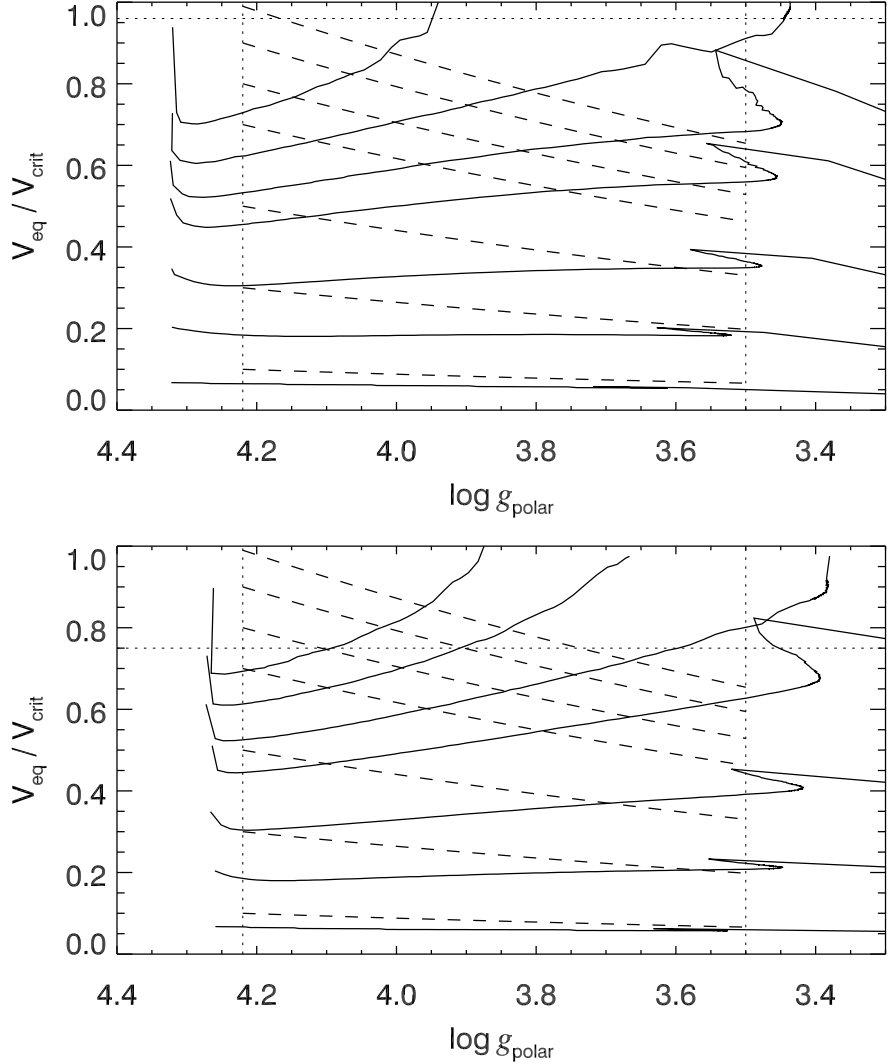}
 \end{center}}
\caption{Evolutionary changes in the EMMB models 
for $M = 3 M_\odot$ ({\it top panel}) and $9 M_\odot$
({\it bottom panel}).  The solid curve lines are for models with the
initial ratios of $\Omega/\Omega_{\rm crit}=0.1$, 0.3, 0.5,
0.7, 0.8, 0.9, and 0.99 from bottom to top, respectively.
The two vertical dotted lines in each panel mark the 
range in evolutionary state for the calculation of the mean velocity.
The horizontal dotted line in each panel marks the rotational
upper limit for normal B stars.  The dashed lines in both panels
are based on our case B models.
}
\label{fig8}
\end{figure}

The EMMB models provide the key data on how stellar rotation evolves 
with time for initial (ZAMS) values of angular rotational rate  
$\Omega/\Omega_{\rm crit}=$ 0.1, 0.3, 0.5, 0.7, 0.8, 0.9, and 0.99.  
Figure~\ref{fig8} shows how the $V_{\rm eq}/V_{\rm crit}$ ratio changes with 
$\log g_{\rm polar}$ in the models for each of these 
starting rotational rates.  We see that the EMMB models predict 
that a sharp drop occurs at the outset for stars with higher initial
$V_{\rm eq}/V_{\rm crit}$ values.  This results from a rapid 
redistribution of angular momentum as the models transform 
from an assumed solid body rotation state at ZAMS into 
a differentially rotating configuration.  Our YMS star sample 
is best compared to the models that have stabilized and 
that are characterized by a mean $\log g_{\rm polar}$ similar
to that of our YMS samples.  Thus, we adopted $\log g_{\rm polar}=4.22$
as the nominal starting point for the model calculations. 
Then the deconvolved distribution of $V_{\rm eq}/V_{\rm crit}$ 
derived from the YMS stars maps into a probability density 
for each of the EMMB tracks shown in Figure~\ref{fig8}.
Note, however, that at this stage of evolution 
($\log g_{\rm polar}=4.22$), the EMMB tracks only sample a range 
from $V_{\rm eq}/V_{\rm crit}=0.0$ to 0.7, and we need
to extrapolate beyond these tracks for the portion of the 
rotational distribution with the fastest spins. 

\placefigure{fig8}     

\begin{figure}
{\begin{center}
 \includegraphics[angle=0,height=17cm]{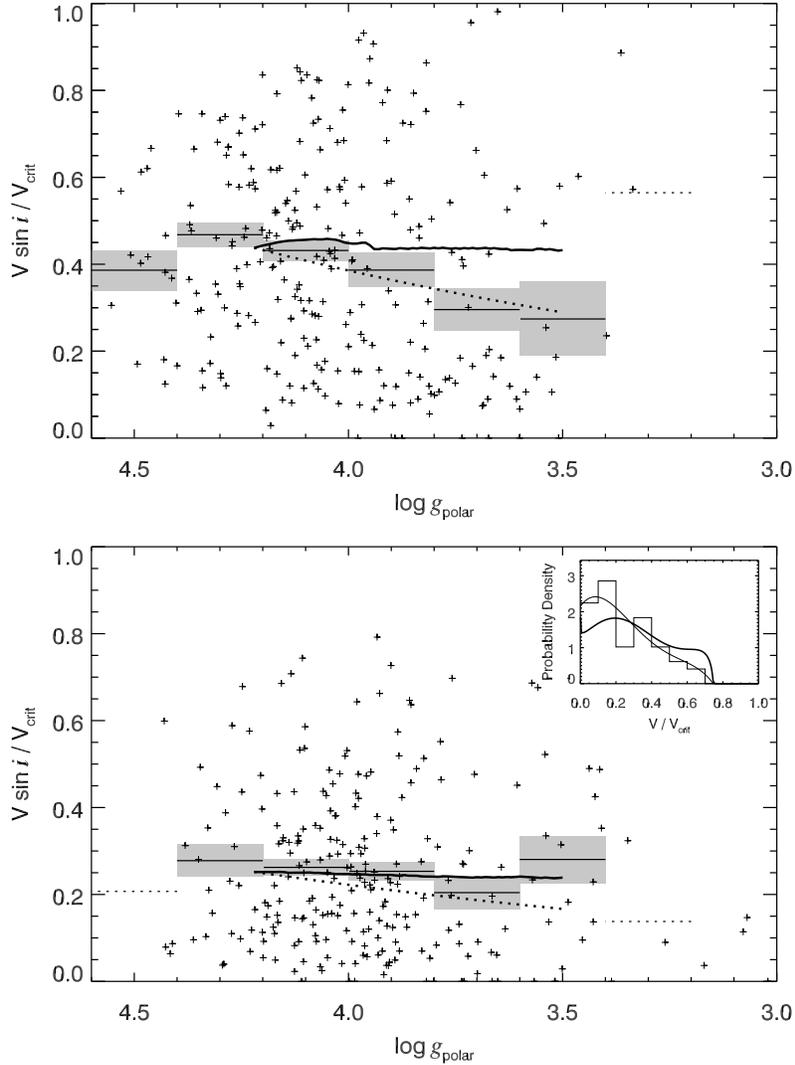}
 \end{center}}
\caption{The $V \sin i/ V_{\rm crit}$ vs. $\log g_{\rm polar}$
plots for B stars in mass range $2 < M/M_\odot < 4$ ({\it top
panel}) and in the range $7 \leq M/M_\odot < 13$ ({\it bottom panel}), 
using the same format as in Fig.~\ref{fig4}.  The thick curves are theoretical 
predictions based on $3 M_\odot$ ({\it top panel}) and
9 $M_\odot$ ({\it bottom panel}) models from \citet{eks08}.
The predictions of the case B 
model are plotted as thick dotted lines.  The embedded plot in
the bottom panel shows the rotation distribution curves for the
YMS stars with $7 \leq M/M_\odot < 13$ (same format as Fig.~\ref{fig5}).
}
\label{fig9}
\end{figure}

We then performed a numerical integration in small steps of 
$\triangle \log g_{\rm polar}$ to find how the probability 
becomes redistributed in $V_{\rm eq}/V_{\rm crit}$ as 
evolution advances.  From inspection of Figure~\ref{fig8}, the 
part of the distribution for the slowly spinning stars 
will be shifted only slightly higher, while the portions 
of the distribution for the fastest rotators will 
be shifted to significantly higher $V_{\rm eq}/V_{\rm crit}$. 
\citet{eks08} argue that these rapidly rotating stars will evolve 
towards the critical limit, where they may become Be stars and 
drain their angular momentum into a circumstellar disk.  
However, we purposely removed the Be stars from our observational 
sample, so we need to truncate the fastest portion of the evolved 
model distributions of $V_{\rm eq}/V_{\rm crit}$ in order to 
make a valid comparison with the observed distributions. 
From our results in \S6 (see Table~7), we estimate that the 
upper limit boundary separating the non-emission line and Be stars
occurs near $V_{\rm eq}/V_{\rm crit} = 0.96$ for a $3 M_\odot$ star
and near $V_{\rm eq}/V_{\rm crit} = 0.75$ for a $9 M_\odot$ star.
These limits are indicated by dotted, horizontal lines in Figure~\ref{fig8}.
Thus, we treated any part of the model $V_{\rm eq}/V_{\rm crit}$ 
distribution that was transformed above these limits as removed
from the sample.  We then determined the mean value of 
$<V_{\rm eq}/V_{\rm crit}>$ from the remaining model distribution at 
each increment of $\log g_{\rm polar}$ (over the range 
$4.22 > \log g_{\rm polar} > 3.5$; indicated by vertical 
dotted lines in Fig.~\ref{fig8}).  Finally, this statistic was  
multiplied by $\pi/4$ (assuming random orientation) to 
obtain a model prediction of $<V \sin i/V_{\rm crit}>$.
These calculated, mean rotational rates are plotted as thick solid lines 
in Figure~\ref{fig9} for the $3 M_\odot$ and $9 M_\odot$ cases, and the predictions 
suggest that the sample means should be almost constant 
over the main sequence lifetime.  This may appear 
somewhat surprising, given the general trend of increasing
$<V_{\rm eq}/V_{\rm crit}>$ for the tracks shown in Figure~\ref{fig8}, 
but recall that the rapid rotators are systematically removed
from the sample distribution once they exceed the nominal Be star 
boundary.  This removal of the high end part of the 
distribution tends to cancel the effect of the small increase 
in $<V_{\rm eq}/V_{\rm crit}>$ for the lower velocity portion
and results in almost no change in the distribution mean value.  

\placefigure{fig9}     

We also show in Figure~\ref{fig9} the observed distributions of 
$V \sin i/V_{\rm crit}$ for the two mass samples.  
The mean values of $V \sin i/V_{\rm crit}$ of the observed stars in
each 0.2 dex bin of $\log g_{\rm polar}$ are plotted as
horizontal bars, and their associated standard deviations of the
mean are shown as gray shaded regions (as in Fig.~\ref{fig4}).  
The differences between the two mass groups are striking. 
The lower mass B stars (top panel of Fig.~\ref{fig9}) rotate much faster
than do the higher mass B stars (lower panel of Fig.~\ref{fig9}) at the beginning
of the MS stage, but the lower mass stars spin down much ``faster''
with respect to $\log g_{\rm polar}$ than do the higher mass stars.  
At the end of the MS stage, both low and high mass B stars have 
roughly the same $<V \sin i/V_{\rm crit}>$.   Our results are 
in qualitative agreement with those of \citet{abt02} and \citet{abt03} 
who examined a similar number of B stars and found that 
between luminosity class V (unevolved) and class III (evolved)
the mean projected rotational velocity $<V \sin i>$ 
changes little for early-B types and drops significantly 
for late-B types. 

A visual appraisal of the predicted and observed evolutionary trends
indicates that the EMMB model prediction makes a good match of
the observational results for the high mass, B star group but 
fails to reproduce the observed spin-down of the lower mass group. 
The evolutionary trends in the EMMB models are directly related to 
their formulation of the angular momentum redistribution in the 
stellar interior caused by mixing.  The EMMB models occupy the 
middle ground between two extreme cases discussed by \citet{oke54}.
In case A of \citet{oke54}, the angular momentum is fully fixed mixed 
within the interior and the star rotates as a solid body.  
As the star evolves with time, it becomes more concentrated 
towards its center (and its moment of inertia gradually
decreases), and consequently the outer layers spin up in terms of 
$V_{\rm eq}/V_{\rm crit}$.  Case B, on the other hand, imagines the
star as a sequence of spherical mass shells in which each 
shell maintains its angular momentum.  Then the surface rotational
velocity will drop in proportion to the increase in radius with 
advanced evolution.  Case A models, with their greater spin up, 
would tend to increase faster with advancing evolution than
the EMMB models shown in Figure~\ref{fig9}, while case B models, with
more spin down, would decrease faster with advancing age.
Since the observations suggest more spin down for the 
$3 M_\odot$ group than predicted by the EMMB model, it is 
worthwhile considering the case B alternative. 

In case B, we suppose that angular momentum is conserved in the 
individual spherical shells of a rotating star.  Then,  
conservation of angular momentum in the outermost layer
(where we measure $V \sin i$) directly leads to $V_{\rm eq} \propto r^{-1}$.
Since $V_{\rm crit} \propto r^{-1/2}$, the ratio will vary as 
$V_{\rm eq}/V_{\rm crit} \propto r^{-1/2} \propto g_{\rm polar}^{1/4}$. 
The predicted trends of spin-down with $\log g_{\rm polar}$ for 
the case B model are plotted as dashed lines in Figure~\ref{fig8}. 
Then we can calculate how the distributions of $V_{\rm eq}/V_{\rm crit}$
will change with advanced evolution according to the case B model 
and determine the mean $<V \sin i/V_{\rm crit}>$ in the same way as we did 
for the EMMB models.  The case B model predictions are plotted as 
thick, dotted lines in Figure~\ref{fig9}.

Surprisingly, the prediction from the simple case B model makes 
a good fit of the spin-down trend observed for the low mass group.
For the high mass group, the fits from the case B and EMMB models
are both reasonably good (except for the lowest gravity bin at 
$\log g_{\rm polar}\sim3.5$ where the EMMB model fit is better).
The small difference in the predictions between the EMMB and
case B models is due to the placement of most of the high mass stars
in the lower $V_{\rm eq}/V_{\rm crit}$ region (bottom inset panel of Fig.~\ref{fig9}) 
where the evolutionary changes of both models are modest (see Fig.~\ref{fig8}).
Note that the case B model ignores any angular momentum loss
from the surface caused by stellar winds.  Such loss is negligible
for the weak winds of the lower mass stars, but inclusion of the 
influence of the stronger winds of the higher mass stars would cause 
a somewhat larger spin-down than illustrated in Figure~\ref{fig9} and might 
result in a more discrepant fit. 

We suspect that the failure of the EMMB models to match the 
faster spin-down we observe in the lower mass stars is related 
to missing physical processes that involve magnetism. 
The transport of internal angular momentum is a complex 
process that involves mixing, magnetic fields, and possibly
gravity waves \citep{tal08,dec09}.  For example, \citet{mae09} 
describe how differential rotation, meridional circulation, and 
mixing can generate a magnetic dynamo in some circumstances. 
The magnetic braking associated with the magnetic field and mass
loss may then lead to a significant spin down of the star. 
We speculate that such magnetic braking is more apparent in the 
lower mass stars because their MS lifetimes are relatively much 
longer and consequently magnetic braking has a greater cumulative 
effect. 


\section{Consequences for Be Star Formation}    

Our results offer some insight about the evolutionary stages 
where we might expect to find the rapid rotation associated
with disk formation in Be stars.  There are three broad 
explanations for the origin of the rapid rotating Be stars: 
1) they were born as rapid rotators; 2) angular momentum 
redistribution with evolution causes their outer layers
to spin up \citep{eks08}; and 3) they are the mass gainers
that were spun up by a past episode of Roche lobe overflow
in an interacting binary system \citep{pol91}.  
We showed in \S5 that most young B stars have rotational 
velocities that are well below the limit for Be star formation 
(compare Fig.~\ref{fig6} for the YMS stars with Fig.~\ref{fig7} for the 
upper limits of non-emission line stars).  Consequently, 
we think that only a small fraction of Be stars were 
born as rapid rotators, and below we will confront the 
expectations from the other two explanations with our results.  

We showed in the last section (see Fig.~\ref{fig9}, lower panel) that 
the EMMB model predictions about evolutionary changes 
in $<V\sin i / V_{\rm crit}>$ are consistent with the 
observed means for the higher mass stars.  The EMMB models
suggest that many of the stars in the upper part of the rotational
distribution will evolve towards critical rotation (Fig.~\ref{fig8}, 
lower panel) and presumably become Be stars. 
\citet{mcs05} found that the Be fraction increases
as high mass B stars evolve.  Their data show
that the Be fraction of early B stars is 4.7\% while that of 
the evolved B stars increases to 7.4\%.  \citet{zor05} also 
found that the Be star fraction increases between the early
to late MS stages in Galactic stars in the mass range 
from 5 to $12 M_\odot$, and \citet{mar06} confirmed 
a similar trend for Be stars in the Large Magellanic Cloud (LMC).
All these results are consistent with the EMMB picture that rapid
rotation develops through the angular momentum redistribution that
occurs as the stars approach the TAMS. 
 
Using the EMMB model shown in the bottom panel of Figure~\ref{fig8}
and our YMS data in the embedded portion of Figure~\ref{fig9}, we can
calculate how the Be fraction might change with time.  
According to the EMMB models, $V_{\rm eq}/V_{\rm crit}$ 
will generally increase with decreasing $\log g_{\rm polar}$,
and once a star reaches the critical limit, it will remain 
there until the completion of core H burning. 
Suppose that a fraction $x$ of the sample are Be stars at the 
outset of the YMS star distribution (i.e., born as rapid rotators). 
Then, we can use the EMMB tracks to determine the number of 
high mass stars that should cross the Be formation limit 
as a function of evolutionary $\log g_{\rm polar}$. 
The calculation indicates that at $\log g_{\rm polar}=4.1$, the Be
fraction is only about $x+3\%$ (formed from all the YMS B stars with
$V_{\rm eq}/V_{\rm crit} > 0.7$, see the embedded plot of Fig.~\ref{fig9}); 
at $\log g_{\rm polar}=3.9$, the Be fraction increases to $x+12\%$ 
(from the YMS B stars with $V_{\rm eq}/V_{\rm crit} > 0.6$); and 
at $\log g_{\rm polar}=3.6$, the Be fraction increases to $x+22\%$
(from the YMS B stars with $V_{\rm eq}/V_{\rm crit} > 0.5$).
This final estimate is in reasonable agreement with the results from 
\citet{mcs05} (Be fraction for evolved B stars of $7.4\%$),  
given that the Be detection efficiency may be as low as 25 -- 50\%
\citep{mcs08}.

However, the situation is different for the lower mass B stars. 
We found that the mean spin rates of the lower mass stars decrease 
faster than predicted by the EMMB model (Fig.~\ref{fig9}, upper panel), 
which probably indicates that these stars evolve to lower 
$V_{\rm eq}/V_{\rm crit}$ (dashed lines in Fig.~\ref{fig8}, upper panel). 
This conclusion is supported by the statistics of late type Be stars.  
If the EMMB models did apply to the lower mass B stars, then 
we can calculate the Be star fraction with time in the 
same way as we did above for the higher mass stars.  
Since the lower mass star distribution has a higher mean velocity,
a larger proportion of the sample are predicted to end up as Be
stars (see Fig.~\ref{fig8}, upper panel).  For example, all the low mass stars  
with $V_{\rm eq}/V_{\rm crit} > 0.6$ at $\log g_{\rm polar}=4.22$
will be transformed into Be stars during the MS stage even 
though the boundary for Be formation is very high.  According to the
top panel of Fig.~\ref{fig6}, about 45\% of low mass YMS stars have
$V_{\rm eq}/V_{\rm crit} > 0.6$, and all of them are predicted to become  
Be stars before they reach the end of the MS stage.  \citet{mcs05}
found that the Be fraction of the late B stars is only about
0.9\%, which is much lower than the prediction from the EMMB model. 
Thus, the low observed Be fraction among the lower mass stars
offers indirect evidence that these stars spin down faster than 
predicted.  

This line of reasoning suggests that lower mass stars spin down 
with time.  If this is also true of low mass Be stars, then 
any such star must be in a state where it obtained its high spin 
very recently, i.e., it is one of the rare newborn stars with 
fast spin or it was spun up through binary mass transfer.   
We note, for example, that Regulus was recently discovered 
to be a spectroscopic binary with a low mass companion 
\citep{gie08} that is probably the white dwarf remnant of 
a prior mass donor \citep{rap09}.  We suspect that this
past mass transfer did spin up Regulus to the critical rate
and that Regulus may have been a Be star in the past. 
However, with the passage of time, it has now spun down 
to a level below the critical level and its rotational rate 
decline will likely continue.  If this scenario for Regulus 
is correct, then we suspect that most of the low mass 
Be stars and other rapid rotators were formed through mass exchange. 
Their binary character may be tested through dedicated 
spectroscopy and radial velocity measurements.  


\section{Conclusions}    

Two recent observing campaigns on cluster and field
B star provide us with new rotational data on about 600 targets.  
Combined with data obtained in our previous surveys, we now have a 
very large sample of B stars that is ideal to investigate many
questions about the rotation of massive stars.  Following
the analysis methods we used in previous study, we derived 
$V \sin i$, $T_{\rm eff}$, and $\log g_{\rm polar}$ 
(considering rotational distortion and gravity darkening)
and estimated the stellar mass and $V_{\rm crit}$
for each star.  We summarize our findings below:

1) The radial velocity measurements allow us to identify the short
period ($P < 14$ d) spectroscopic binaries (SB).  The fraction 
of SBs in our field and cluster samples are similar.  
Considering the efficiency of detection, we estimate that about 
19\% of B stars are short period SBs.  This result is consistent
with many previous studies of binary frequency of B stars \citep{wol78,abt90}.

2) The mean $V \sin i$ for the field sample (441 stars) is
slower than that for the cluster sample (557 stars), confirming
results from previous studies.  By comparing
stars with similar evolutionary status, we find that the stars
in these two samples have similar rotational properties when 
plotted as a function of $\log g_{\rm polar}$.  The mean rotation 
rate of the field B sample is slower than that for the cluster
sample because the former contains a larger fraction of evolved stars.

3) The rotation distribution curves based on young stars 
with $\log g_{\rm polar} > 4.15$ suggest that
massive stars are born at various rotation rates, including some
very slow rotators ($V_{\rm eq}/V_{\rm crit} <0.1$).  
The mass dependence suggests that higher mass B stars may
preferentially experience angular momentum  loss processes
during and after formation. 
The low mass stars are born with more rapid rotators than
high mass stars if the stellar rotation rate is evaluated
by $V_{\rm eq}/V_{\rm crit}$.  

4) The statistics based on the normal B (non-Be) stars in our sample
indicates that low mass B stars ($< 4 M_\odot$) may require a high threshold
of $V_{\rm eq}/V_{\rm crit} > 0.96$ to become Be stars.  As
stellar mass increases, this threshold decreases, dropping to 
0.64 for B stars with $M> 8.6 M_\odot$.  This implies that 
the mass loss processes leading to disk formation may be very different 
for low and high mass Be stars. 

5) Comparing with modern evolutionary models of rotating stars (for 3 and 
$9 M_\odot$ from \citealt{eks08}), the apparent evolutionary trends of 
$<V\sin i /V_{\rm crit}>$  are in good agreement for the 
high mass B stars, but the data for low mass B stars 
shows a more pronounced spin-down trend than predicted.  

6) Predictions for the fractions of rapid rotators and Be stars 
produced by the redistribution of angular momentum with evolution
agree with observations for the higher mass B stars but vastly 
overestimate the Be population for the lower mass stars. 
The greater than expected spin-down of the lower mass stars 
explains this discrepancy and suggests that most of the 
low mass Be stars were probably spun up recently.  


\acknowledgments
We thank an anonymous referee for his/her careful reading and
valuable comments that improved our paper in many ways.
We thank George Will, Dianne Harmer, Daryl Willmarth, and the KPNO
staff for their assistance in making these observations successful.
We thank Yale University for access to the WIYN telescope through
their local TAC.  This material is based upon work supported
by the National Science Foundation under Grant No.~AST-0606861 (DRG).
WJH thanks George Wallerstein and the Kenilworth Fund of the
New York Community Trust for partial financial support of this study.
WJH is also very grateful for partial finance support from NSF grant
AST-0507219 to Judith G. Cohen.  MVM is grateful for support
from NSF grant AST-0401460 as well as Lehigh University. This
research has made use of the SIMBAD database, operated at CDS,
Strasbourg, France.  This research has made use of the WEBDA
database, operated at the Institute for Astronomy of the University
of Vienna.  We have made use of the images from the Digitized Sky
Surveys, which were produced at the Space Telescope Science
Institute under U.S. Government grant NAG W-2166.




\clearpage

\begin{deluxetable}{lcc}
\tablewidth{0pc}
\tablecaption{Open Clusters Observed
\label{tab1}}
\tablehead{
\colhead{ } &
\colhead{$\log$ Age } &
\colhead{Number of}\\
\colhead{Name } &
\colhead{(y) } &
\colhead{Stars Observed}}
\startdata
 IC 4996  &  6.95   &  16  \\
 NGC 581  &  7.34   &  22  \\
 NGC 869  &  7.07   &  35  \\
 NGC 884  &  7.03   &  34  \\
 NGC 1893 &  7.03   &  22  \\
 NGC 1960 &  7.47   &  27  \\
 NGC 6871 &  6.96   &  25  \\
 NGC 7380 &  7.08   &  22  \\
 NGC 7654 &  7.76   &  31  \\
\enddata
\end{deluxetable}


\begin{deluxetable}{rrr lcc ccc ccc}
\tabletypesize{\footnotesize} 
\rotate
\tablewidth{0pc}
\tablecaption{Spectroscopic Binaries Common to the Field B Star Sample and SB9\tablenotemark{a}
\label{tab2}}
\tablehead{
\multispan{7}{\hfil SB9 \hfil} &
\colhead{ } &
\multispan{3}{\hfil Our Measurements \hfil} &
\colhead{ } \\
\cline{1-7}
\cline{9-11}
\colhead{SB9}               &
\colhead{ }                 &
\colhead{$P$}               &
\colhead{ }                 &
\colhead{$K_1$}             &
\colhead{$K_2$}             &
\colhead{$\gamma$}          &
\colhead{ }                 &
\colhead{$V_{\rm r}$(N1)}   &
\colhead{$V_{\rm r}$(N2)}   &
\colhead{$|\Delta V_{\rm r}|$} &
\colhead{ }                 \\
\colhead{Index}             &
\colhead{ HD }              &
\colhead{(d)}               &
\colhead{ $e$ }             &
\colhead{(km s$^{-1}$)}     &
\colhead{(km s$^{-1}$)}     &
\colhead{(km s$^{-1}$)}     &   
\colhead{ }                 &        
\colhead{(km s$^{-1}$)}     &        
\colhead{(km s$^{-1}$)}     &        
\colhead{(km s$^{-1}$)}     &         
\colhead{ Detected} }
\startdata
 131 &  16219 &   2.1 & 0.05 &\phn    21 & \nodata &\phs    12 & &\phs\phn 4 &\phs    28 &    24 & y \\
1353 & 209961 &   2.2 & 0.03 &       122 & \nodata &     $-$13 & &     $-$14 &     $-$19 &\phn 5 & n \\
 189 &  23466 &   2.4 & 0.0~ &\phn    23 & \nodata &\phs    17 & &\phn  $-$7 &\phs    33 &    39 & y \\
 484 &  65041 &   2.8 & 0.30 &\phn    34 & \nodata &     $-$13 & &\phs    26 &     $-$29 &    55 & y \\
1423 & 218407 &   3.3 & 0.24 &\phn    86 & \nodata &     $-$10 & &\phn  $-$2 &     $-$69 &    67 & y \\
 146 &  17543 &   3.9 & 0.04 &\phn    25 & \nodata &\phs\phn 8 & &\phn\phs 7 &     $-$11 &    17 & y \\
 313 &  34762 &   5.4 & 0.08 &\phn    27 & \nodata &\phs\phn 6 & &\phn  $-$1 &     $-$17 &    16 & y \\
1424 & 218440 &   7.3 & 0.38 &\phn    88 &   147   &\phn  $-$5 & &     $-$22 &     $-$10 &    13 & n \\
 394 &  44172 &   8.2 & 0.0~ &\phn    25 &\phn 48  &\phn  $-$6 & &     $-$24 &     $-$17 &\phn 7 & n \\
1409 & 216916 &  12.1 & 0.05 &\phn    24 & \nodata &     $-$12 & &     $-$12 &\phs\phn 3 &    15 & y \\
 375 &  41040 &  14.6 & 0.39 &\phn    35 &\phn 39  &\phs    13 & &\phs    11 &\phs\phn 9 &\phn 2 & n \\
  17 &   1976 &  25.4 & 0.12 &\phn    24 & \nodata &     $-$10 & &\phs    12 &\phs\phn 2 &    10 & n \\
  44 &   4382 &  33.8 & 0.41 &\phn    16 & \nodata &\phn  $-$4 & &\phn\phs 5 &\phn  $-$8 &    13 & y \\
2401 &  20315 &  36.5 & 0.3~ &\phn    20 & \nodata &\phs\phn 4 & &\phn  $-$9 &\phn  $-$9 &\phn 0 & n \\
 297 &  32990 &  58.3 & 0.19 &\phn    37 & \nodata &\phs    16 & &\phs    36 &\phs    34 &\phn 2 & n \\
  19 &   2054 &  48.3 & 0.38 &\phn    30 & \nodata &\phs\phn 3 & &     $-$14 &\phn  $-$7 &\phn 7 & n \\
  95 &  11529 &  69.9 & 0.30 &\phn    30 & \nodata &     $-$25 & &\phn\phs 7 &\phn\phs 4 &\phn 3 & n \\
  29 &   3322 & 399.6 & 0.57 &\phn\phn 8 & \nodata &\phs\phn 4 & &\phs    11 &\phn\phs 8 &\phn 3 & n \\
  66 &   7374 & 800.9 & 0.31 &\phn\phn 4 & \nodata &     $-$15 & &     $-$21 &     $-$16 &\phn 4 & n \\
  36 &   3901 & 940.2 & 0.4~ &\phn    12 & \nodata &     $-$11 & &\phn  $-$4 &\phn  $-$2 &\phn 2 & n \\
\enddata
\tablenotetext{a}{The Ninth Catalogue of Spectroscopic Binary Orbits \citep{pou04}.}
\end{deluxetable}


\begin{deluxetable}{lcc ccc ccc ccc cc}
\tabletypesize{\scriptsize}
\rotate
\tablewidth{0pc}
\tablecaption{Stellar Parameters of New Cluster B Stars
\label{tab3}}
\tablehead{
\colhead{Cluster } &
\colhead{WEBDA } &
\colhead{$V_{\rm r}$(N1)\tablenotemark{a}} &
\colhead{$V_{\rm r}$(N2)\tablenotemark{a}} &
\colhead{$V \sin i$} &
\colhead{$\Delta V \sin i$} &
\colhead{$T_{\rm eff}$} &
\colhead{$\Delta T_{\rm eff}$} &
\colhead{ } &
\colhead{ } &
\colhead{ } &
\colhead{Mass} &
\colhead{$V_{\rm crit}$ } &
\colhead{ } \\
\colhead{Name} &
\colhead{ID } &
\colhead{(km~s$^{-1}$)} &
\colhead{(km~s$^{-1}$)} &
\colhead{(km~s$^{-1}$)} &
\colhead{(km~s$^{-1}$)} &
\colhead{(K)} &
\colhead{(K)} &
\colhead{$\log g$} &
\colhead{$\Delta \log g$} &  
\colhead{$\log g_{\rm polar}$} &
\colhead{($M_\odot$)} &
\colhead{(km~s$^{-1}$)} &
\colhead{Comment}}
\startdata
IC4996  &   4 & ~$-$6 & ~$-$6 & ~25 & 12 & 10508 & ~50 & 4.183 & 0.022 & 4.193 & ~2.6 &  392 & \nodata \\
IC4996  &   6 & $-$20 & $-$26 & 156 & ~7 & 10270 & ~50 & 3.903 & 0.010 & 4.032 & ~2.7 &  361 & \nodata \\
IC4996  &   8 & ~$-$5 & $-$13 & ~64 & ~9 & 26014 & 550 & 4.262 & 0.062 & 4.273 & 10.4 &  582 & \nodata \\
\nodata & \nodata & \nodata & \nodata & \nodata & \nodata & \nodata & \nodata & \nodata & \nodata & \nodata & \nodata & \nodata & \nodata \\
\enddata
\tablenotetext{a}{The HJD times for $V_{\rm r}$ are given in the online version
of this table.}
\end{deluxetable}


\begin{deluxetable}{rrc ccc ccc ccc ccc c}
\tabletypesize{\scriptsize}
\rotate
\tablewidth{0pc}
\tablecaption{Stellar Parameters of New Field B Stars
\label{tab4}}
\tablehead{
\colhead{  } &   
\colhead{  } &
\colhead{HJD(N1) } &
\colhead{$V_{\rm r}$(N1) } &
\colhead{HJD(N2) } &
\colhead{$V_{\rm r}$(N2) } &
\colhead{$V \sin i$} &
\colhead{$\Delta V \sin i$} &
\colhead{$T_{\rm eff}$} &
\colhead{$\Delta T_{\rm eff}$} &
\colhead{} &
\colhead{} &
\colhead{} &  
\colhead{Mass} & 
\colhead{$V_{\rm crit}$} &
\colhead{} \\
\colhead{SAO} &
\colhead{HD } &
\colhead{2454000$+$ } &
\colhead{(km~s$^{-1}$)} &
\colhead{2454000$+$ } &
\colhead{(km~s$^{-1}$)} &
\colhead{(km~s$^{-1}$)} &
\colhead{(km~s$^{-1}$)} &
\colhead{(K)} &
\colhead{(K)} & 
\colhead{$\log g$} &
\colhead{$\Delta \log g$} &
\colhead{$\log g_{\rm polar}$} &
\colhead{($M_\odot$)} &
\colhead{(km~s$^{-1}$)} &
\colhead{Comment}}
\startdata
4080  &   1359 & 790.669  & 1.5 & 791.659 & $-$2.3 & 203 & 12 & 12000 & ~50 &  3.834 & 0.012 &  4.007 &  3.4 & 376 & \nodata \\
4165  &   3366 & 790.671  & 2.8 & 791.662 & $-$5.1 & ~~3 & 21 & 17729 & 250 &  3.886 & 0.027 &  3.887 &  6.3 & 411 & \nodata \\
4226  &   4382 & 790.675  & 5.0 & 791.664 & $-$8.3 & ~72 & 31 & 12612 & 150 &  3.299 & 0.029 &  3.393 &  4.9 & 290 & \nodata \\
\nodata & \nodata & \nodata & \nodata & \nodata & \nodata & \nodata & \nodata & \nodata & \nodata & \nodata & \nodata & \nodata & \nodata & \nodata & \nodata \\
\enddata
\end{deluxetable}


\begin{deluxetable}{lccc}
\tablewidth{0pc}
\tablecaption{Summary of the B Star Samples
\label{tab5}}
\tablehead{  
\colhead{Sample } &
\colhead{$N$($V_{\rm r}$ const.) } &
\colhead{$N$($V_{\rm r}$ var.) } &  
\colhead{$N$(Total)}}
\startdata
New Field                       & 332                  &     43    &    375  \\
New Cluster                     & 205                  &     27    &    234\tablenotemark{a}  \\
Prior Field\tablenotemark{b}    & \phn 91              &     13    &    104  \\
Prior Cluster\tablenotemark{c}  & 327                  &     78    &    433\tablenotemark{a}  \\
\enddata
\tablenotetext{a}{Includes some cases where the $V_{\rm r}$ variation status 
is unknown since only one measurement is available.}
\tablenotetext{b}{\citet{hua08}}
\tablenotetext{c}{\citet{hua06b}}
\end{deluxetable}

\clearpage

\begin{deluxetable}{ccc ccc ccc}
\tabletypesize{\footnotesize}
\tablewidth{0pc}
\tablecaption{Rotational Data for the Field and Cluster B Stars with
$6 < M/M_\odot < 12$
\label{tab6}}
\tablehead{
\colhead{ } &
\colhead{ } &
\multispan{3}{\hfil Field \hfil} &
\colhead{ } &
\multispan{3}{\hfil Cluster \hfil} \\
\cline{3-5}
\cline{7-9}
\colhead{ Range of } &
\colhead{ } &
\colhead{ No. of} &
\colhead{ $<V\sin i>$ } &
\colhead{$\sigma_\mu$} &
\colhead{ } &
\colhead{ No. of} &
\colhead{ $<V\sin i>$ } &
\colhead{$\sigma_\mu$} \\
\colhead{$\log g_{\rm polar}$} &
\colhead{ } &
\colhead{ Stars } &
\colhead{ (km s$^{-1}$) } &
\colhead{(km s$^{-1}$)} & 
\colhead{ } &
\colhead{ Stars } &
\colhead{ (km s$^{-1}$) } &
\colhead{ (km s$^{-1}$) }
 }
\startdata
  3.2$\sim$3.4 & & \phn 5 & \phn 26.2  &  25.8 & & \phn\phn 1 & \phn\phn 0.0 &  \nodata  \\
  3.4$\sim$3.6 & &     17 & \phn 76.5  &  18.6 & & \phn    10 &        102.1 &     27.9  \\
  3.6$\sim$3.8 & &     19 & \phn 77.4  &  19.8 & & \phn    17 & \phn    95.5 &     17.6  \\
  3.8$\sim$4.0 & &     18 &     107.7  &  26.5 & & \phn    82 &        115.1 & \phn 9.6  \\
  4.0$\sim$4.4 & &     15 &     163.3  &  34.5 & &        136 &        135.1 & \phn 9.1  \\
\enddata

\end{deluxetable}

\clearpage

\begin{deluxetable}{ccc ccc ccc}
\tabletypesize{\footnotesize}
\rotate
\tablewidth{0pc}
\tablecaption{Statistical Results for the Fastest Rotating Normal B Stars
\label{tab7}}
\tablehead{
\colhead{Mass Range } &
\colhead{ } &
\multispan{3}{\hfil From~$V\sin i/V_{\rm crit}$ distribution \hfil} &
\colhead{ } &
\multispan{2}{\hfil From~$V_{\rm eq}/V_{\rm crit}$ distribution \hfil} &
\colhead{ $(V_{\rm eq}/V_{\rm crit})_{\rm upper~limit}$} \\
\cline{3-5}
\cline{7-8}
\colhead{($M_\odot$) } &
\colhead{$N$ } &
\colhead{($V\sin i/V_{\rm crit}$)$_{\rm max}$ } &
\colhead{$<V\sin i/V_{\rm crit}>_{\rm 1\%}$} &
\colhead{$<V\sin i/V_{\rm crit}>_{\rm 3\%}$} &
\colhead{ } &
\colhead{$(V_{\rm eq}/V_{\rm crit})_{\rm 4\%}$} &
\colhead{$(V_{\rm eq}/V_{\rm crit})_{\rm 0.2\%}$ } &
\colhead{ for non-Be stars}\\
\colhead{(1)} & 
\colhead{(2)} & 
\colhead{(3)} & 
\colhead{(4)} & 
\colhead{(5)} & 
\colhead{ } & 
\colhead{(6)} & 
\colhead{(7)} &
\colhead{(8)} 
 }
\startdata
2.2$\sim$4.0  &      255 &     0.98  &    0.96  &   0.91 & & 0.93 & 0.99 & 0.99$^{+0.01}_{-0.07}$ \\
4.0$\sim$5.4  &      225 &     0.91  &    0.88  &   0.84 & & 0.87 & 0.94 & 0.94$^{+0.06}_{-0.08}$ \\
5.4$\sim$8.6  &      255 &     0.82  &    0.81  &   0.76 & & 0.79 & 0.86 & 0.86$^{+0.08}_{-0.08}$ \\
8.6$\sim$max  &      159 &     0.59  &    0.58  &   0.55 & & 0.57 & 0.63 & 0.63$^{+0.07}_{-0.07}$ \\
all           &      894 &     0.98  &    0.91  &   0.85 & & 0.88 & 0.95 & \nodata \\
\enddata

\end{deluxetable}

\clearpage

\begin{deluxetable}{lcc c}
\tabletypesize{\footnotesize}
\tablewidth{0pc}
\tablecaption{The Highest $V\sin i/V_{\rm crit}$ Stars
\label{tab8}}
\tablehead{
\colhead{Star\tablenotemark{a} } &
\colhead{ Mass ($M_\odot$)} &
\colhead{ $V\sin i/V_{\rm crit}$} &
\colhead{ $\log g_{\rm polar}$ } }
\startdata

HD236849     & ~3.6  & 0.98 &  3.652 \\
NGC7160-940  & ~3.8  & 0.96 &  3.714 \\
HD26793      & ~3.4  & 0.93 &  3.965 \\
HD12352      & ~2.9  & 0.92 &  3.976 \\
HD8053       & ~3.5  & 0.91 &  3.942 \\
HD9766       & ~3.9  & 0.89 &  3.364 \\
HD87901      & ~3.5  & 0.87 &  3.950 \\
HD53744      & ~3.3  & 0.86 &  3.818 \\
\cline{1-4}
HD52860      & ~4.0  & 0.91 &  3.380 \\
NGC869-1417  & ~4.3  & 0.86 &  4.175 \\
NGC7654-1174 & ~4.2  & 0.85 &  3.904 \\
NGC7654-1272 & ~4.1  & 0.84 &  3.949 \\
HD5882       & ~4.6  & 0.82 &  4.131 \\
HD2626       & ~4.3  & 0.80 &  3.753 \\
NGC1960-91   & ~4.2  & 0.79 &  4.195 \\
\cline{1-4}
NGC1893-256  & ~5.5  & 0.82 &  4.267 \\
NGC6871-14   & ~6.2  & 0.81 &  4.170 \\
NGC3293-28   & ~8.3  & 0.79 &  3.933 \\
NGC4755-139  & ~5.6  & 0.75 &  4.266 \\
NGC3293-29   & ~7.2  & 0.74 &  4.108 \\
NGC6871-16   & ~6.1  & 0.74 &  4.151 \\
NGC884-2255  & ~7.8  & 0.73 &  3.901 \\
NGC884-2622  & ~6.3  & 0.73 &  4.212 \\
\cline{1-4}
NGC884-2190  & ~9.8 & 0.59 &  4.102 \\
HD52918      & 10.1 & 0.57 &  3.885 \\
IC2944-122   & ~8.7 & 0.54 &  4.103 \\
HD184915     & 13.8 & 0.53 &  3.791 \\
NGC869-1141  & ~9.7 & 0.51 &  3.824 \\
\enddata
\tablenotetext{a}{The name of a cluster star has
two parts, the cluster name and the WEBDA ID.}
\end{deluxetable}

\clearpage

\clearpage

\clearpage

\clearpage

\clearpage

\clearpage
\begin{thebibliography}{}
\bibitem[Abt et al.(1990)]{abt90}
         Abt, H. A., Gomez, A. E., \&, Levy, S. G. 1990, \apjs, 74, 551
\bibitem[Abt(2003)]{abt03}
         Abt, H. A. 2003, ApJ, 582, 420
\bibitem[Abt(2009)]{abt09}
         Abt, H. A. 2009, PASP, 121, 1307
\bibitem[Abt et al.(2002)Abt, Levato, \& Grosso]{abt02}
         Abt, H. A, Levato, H., \& Grosso, M. 2002, \apj, 573, 359
\bibitem[Chauville et al.(2001)]{cha01}
         Chauville, J., Zorec, J., Ballereau, D., Morrell, N., Cidale, L., \& Garcia, A. 2001, \aap, 378, 861
\bibitem[Cranmer(2005)]{cra05}
         Cranmer, S. R. 2005, \apj, 634, 585
\bibitem[Cranmer(2009)]{cra09}
         Cranmer, S. R. 2009, \apj, 701, 396
\bibitem[Decressin et al.(2009)]{dec09}
         Decressin, T., Mathis, S., Palacios, A., Siess, L., Talon, S., Charbonnel, C., \& 
         Zahn, J.-P. 2009, \aap, 495, 271
\bibitem[Domiciano de Souza et al.(2003)]{dom03}
         Domiciano de Souza, A., et al.\ 2003, \aap, 407, 47
\bibitem[Ekstr\"om et al.(2008)]{eks08}
         Ekstr\"om, S., Meynet, G., Maeder, A., \& Barblan, F. 2008, \aap, 478, 467 [EMMB]
\bibitem[Fr\'{e}mat et al.(2005)]{fre05}
         Fr\'{e}mat, Y., Zorec, J., Hubert, A.-M., \& Floquet, M. 2005, \aap, 440, 305
\bibitem[Gies et al.(2008)]{gie08}
         Gies, D. R., et al. 2008, \apjl, 682, L117
\bibitem[Heger \& Langer(2000)]{heg00}
         Heger, A., \& Langer, N. 2000, \apj, 544, 1016
\bibitem[Howarth(2007)]{how07}
         Howarth, I. D. 2007, in Active OB-Stars: Laboratories for Stellar and 
         Circumstellar Physics (ASP Conf. Ser. 361), ed. S. \v{S}tefl, S. P. Owocki, \& 
         A. T. Okazaki (San Francisco: ASP), 15
\bibitem[Huang \& Gies(2006a)]{hua06a}
         Huang, W., \& Gies, D. R. 2006a, \apj, 648, 580
\bibitem[Huang \& Gies(2006b)]{hua06b}
         Huang, W., \& Gies, D. R. 2006b, \apj, 648, 591
\bibitem[Huang \& Gies(2008)]{hua08}
         Huang, W., \& Gies, D. R. 2008, \apj, 683, 1045
\bibitem[Hunter et al.(2009)]{hun09}
         Hunter, I., et al. 2009, \aap, 496, 841
\bibitem[Kharchenko et al.(2005)]{kha05}
         Kharchenko, N. V., Piskunov, A. E., R\"{o}ser, S., Schilbach, E., \& Scholz, R.-D.
         2005, \aap, 438, 1163
\bibitem[Lanz \& Hubeny(2007)]{lan07}
         Lanz, T., \& Hubeny, I. 2007, \apjs, 169, 83
\bibitem[Larson(2007)]{lar07}
         Larson, R. B. 2007, Rep.\ Prog.\ Phys., 70, 337
\bibitem[Larson(2010)]{lar09}
         Larson, R. B. 2010, Rep.\ Prog.\ Phys., 73, 14901
\bibitem[Lucy(1974)]{luc74}
         Lucy, L. B. 1974, \aj, 79, 745
\bibitem[Maeder et al.(2009)]{mae09}
         Maeder, A., Meynet, G., Georgy, C., \& Ekstr\"{o}m, S. 2009,
         in Cosmic Magnetic Fields: From Planets, to Stars and Galaxies, 
         IAU Symp. 259, Proc. IAU, 4, 
         ed. K. G. Strassmeier, A. G. Kosovichev, \& J. E. Beckman
         (Cambridge: Cambridge Univ. Press), 311
\bibitem[Martayan et al.(2006)]{mar06}
         Martayan, C., Fr\'{e}mat, Y., Hubert, A.-M., Floquet, M., Zorec, J., \& 
         Neiner, C. 2006, \aap, 452, 273
\bibitem[Mason et al.(2009)]{mas09}
         Mason, B. D., Hartkopf, W. I., Gies, D. R., Henry, T. J., \& Helsel, J. W.
         2009, \aj, 137, 3358
\bibitem[McAlister et al.(2005)]{mca05}
         McAlister, H. A., et al. 2005, \apj, 628, 439
\bibitem[McSwain \& Gies(2005)]{mcs05}
         McSwain, M. V., \& Gies, D. R. 2005, \apjs, 161, 118
\bibitem[McSwain et al.(2008)]{mcs08}
         McSwain, M. V., Huang, W., Gies, D. R., Grundstrom, E. D., \& Townsend, R. H. D. 2008, \apj, 672, 590
\bibitem[McSwain et al.(2009)]{mcs09}
         McSwain, M. V., Huang, W., \& Gies, D. R. 2009, \apj, 700, 1216
\bibitem[Meynet \& Maeder(2000)]{mey00}
         Meynet, G., \& Maeder, A. 2000, \aap, 361, 101
\bibitem[Oke \& Greenstein(1954)]{oke54}
         Oke, J. B., \& Greenstein, J. L. 1954, ApJ, 120, 384
\bibitem[Piskunov et al.(2008)]{pis08}
         Piskunov, A. E., Schilbach, E., Kharchenko, N. V., R\"{o}ser, S., \& Scholz, R.-D.
         2008, \aap, 477, 165
\bibitem[Pols et al.(1991)]{pol91}
         Pols, O. R., Cot\'{e}, J., Waters, L. B. F. M., \& Heise, J. 1991, \aap, 241, 419
\bibitem[Porter(1996)]{por96}
         Porter, J. M. 1996, \mnras, 280, L31
\bibitem[Porter \& Rivinius(2003)]{por03}
         Porter, J. M., \& Rivinius, T. 2003, \pasp, 115, 1153
\bibitem[Pourbaix et al.(2004)]{pou04}
         Pourbaix, D., et al.\ 2004, \aap, 424, 727
\bibitem[Rappaport et al.(2009)]{rap09}
         Rappaport, S., Podsiadlowski, Ph., \& Horev, I. 2009, \apj, 698, 666
\bibitem[Rivinius et al.(2003)]{riv03}
         Rivinius, Th., Baade, D., \& \v{S}tefl, S. 2003, \aap, 411, 229
\bibitem[Schaller et al.(1992)]{sch92}
         Schaller, G., Schaerer, D., Meynet, G., \& Maeder, A. 1992, \aaps, 96, 269
\bibitem[Strom et al.(2005)Strom, Wolff, \& Dror]{str05}
         Strom, S. E., Wolff, S. C., \& Dror, D. H. A. 2005, \aj, 129, 809
\bibitem[Talon(2008)]{tal08}
         Talon, S. 2008, in Stellar Nucleosynthesis: 50 years after B$^2$FH, 
         ed. C. Charbonnel \& J.-P. Zahn, EAS Publ. Ser., 32, 81
\bibitem[Townsend et al.(2004)Townsend, Owocki, \& Howarth]{tow04}
         Townsend, R. H. D., Owocki, S. P., \& Howarth, I. D.
         2004, \mnras, 350, 189
\bibitem[Vinicius et al.(2006)]{vin06}
         Vinicius, M. M. F., Zorec, J., Leister, N. V., \& Levenhagen, R. S. 2006, \aap, 446, 643
\bibitem[Wolff (1978)]{wol78}
         Wolff, S. C. 1978, \apj, 222, 556
\bibitem[Wolff et al.(2004)]{wol04}
         Wolff, S. C., Strom, S. E., \& Hillenbrand, L. A. 2004, \apj, 601, 979
\bibitem[Wolff et al.(2007)]{wol07}
         Wolff, S. C., Strom, S. E., Dror, D., \& Venn, K. 2007,  \aj, 133, 1092
\bibitem[Wolff et al.(2008)]{wol08}
         Wolff, S. C., Strom, S. E., Cunha, K., Daflon, S., Olsen, K., \& Dror, D. 2008, \aj, 136, 1049
\bibitem[Yudin(2001)]{yud01}
         Yudin, R. V. 2001, \aap, 368, 912
\bibitem[Zorec et al.(2005)]{zor05}
         Zorec, J., Fr\'{e}mat, Y., \& Cidale, L. 2005, \aap, 441, 235
\end{thebibliography}
\end{document}